\documentclass[11pt]{article}
\usepackage[T1]{fontenc}
\usepackage[letterpaper, margin=1in]{geometry} \usepackage{amsmath, amssymb, amsthm}
\usepackage{bbm}
\usepackage{graphicx}
\usepackage{tabularx}
\usepackage{booktabs}
\usepackage{url}
\usepackage{hyperref}
\usepackage[capitalise]{cleveref}
\usepackage{algorithm}
\usepackage[noend]{algpseudocode}
\usepackage{subcaption}
\usepackage{thmtools}
\usepackage{thm-restate}
\usepackage{mathrsfs}
\usepackage{xcolor}
\usepackage{tikz}
\usetikzlibrary{arrows.meta,backgrounds,decorations.pathreplacing,fit}

\newtheorem{theorem}{Theorem}[section]
\newtheorem{corollary}[theorem]{Corollary}
\newtheorem{lemma}[theorem]{Lemma}

\newtheorem{remark}[theorem]{Remark}

\newenvironment{proof-sketch}{\noindent{\bf Sketch of Proof}\hspace*{1em}}{\qed\bigskip}
\newenvironment{proof-idea}{\noindent{\bf Proof Idea}\hspace*{1em}}{\qed\bigskip}
\newenvironment{proof-of-lemma}[1]{\noindent{\bf Proof of Lemma #1}\hspace*{1em}}{\qed\bigskip}
\newenvironment{proof-attempt}{\noindent{\bf Proof Attempt}\hspace*{1em}}{\qed\bigskip}

\makeatletter
\def\fnum@figure{{\bf Figure \thefigure}}
\def\fnum@table{{\bf Table \thetable}}
\long\def\@mycaption#1[#2]#3{\addcontentsline{\csname
  ext@#1\endcsname}{#1}{\protect\numberline{\csname
  the#1\endcsname}{\ignorespaces #2}}\par
  \begingroup
    \@parboxrestore
    \small
    \@makecaption{\csname fnum@#1\endcsname}{\ignorespaces #3}\par
  \endgroup}
\def\mycaption{\refstepcounter\@captype \@dblarg{\@mycaption\@captype}}
\makeatother

\newcommand{\mathify}[1]{\ifmmode{#1}\else\mbox{$#1$}\fi}
\newcommand{\bigO}O

\newcommand{\remove}[1]{}
\newcommand{\ignore}[1]{}

\def\Z{\mathbb{Z}}

\newcommand{\mc}{\mathcal}

\newcommand{\ceil}[1]{\left\lceil #1 \right\rceil}
\newcommand{\floor}[1]{\left\lfloor #1 \right\rfloor}

\def\poly{{\rm poly}}

\usepackage{enumitem}
\hypersetup{
  colorlinks=true,
  linkcolor=black,
  filecolor=blue,
  urlcolor=blue,
  citecolor=blue,
  pdfpagemode=FullScreen,
}
\usepackage[sortcites, natbib=true, url=false, style=alphabetic, doi=false, isbn=false, maxalphanames=4, maxbibnames=9]{biblatex}
\AtEveryBibitem{
  \clearlist{language}
  \clearfield{location}
}
\newcommand{\E}{\mathbb E}

\newcommand{\ALG}{\mathrm{ALG}}
\newcommand{\OPT}{\mathrm{OPT}}

\title{Collective Tree Exploration with Predictions}
\newcommand{\Fcal}{\mathcal{F}}

\newcommand{\Lcal}{\mathcal{L}}

\newcommand{\Scal}{\mathcal{S}}

\newcommand{\Nbb}{\mathbb{N}}

\usepackage{bbm}

\usetikzlibrary{arrows.meta}

\usepackage[textsize=tiny]{todonotes}

\newcommand{\ch}{\mathrm{ch}}

\begin{document}
\author{Romain Cosson\thanks{Computer Science Department, Courant Institute, New York University. Email: \texttt{romain.cosson@nyu.edu}.} \and Yichen Huang\thanks{John A. Paulson School of Engineering and Applied Science, Harvard University. Supported in part by NSF grant CNS-2107078. Email: \texttt{yichenhuang@g.harvard.edu}.} \and Michael Mitzenmacher\thanks{John A. Paulson School of Engineering and Applied Science, Harvard University. Supported in part by NSF grant CNS-2107078. Email: \texttt{michaelm@eecs.harvard.edu}.}\and Ali Vakilian\thanks{Rutgers University. Email: \texttt{a.vakilian@rutgers.edu}. Supported in part by NSF CAREER award CAIF-2638918.}}
\vspace{10cm}
\date{}

\maketitle

\begin{abstract}
We study collective tree exploration with subtree-size predictions. In this problem, a team of $k$ mobile agents needs to explore an unknown rooted tree as fast as possible, and whenever a new vertex $u$ is discovered, the team receives a prediction $\hat n_u$ of the number of nodes $n_u$ in the subtree rooted at $u$. 
We provide algorithmic guarantees under two error measures: the maximum error $E=\max_u|\widehat n_u-n_u|$ and the multiplicative distortion $\mu = (\max_u\widehat n_u/n_u)(\max_u n_u/\widehat n_u)$. For the maximum error, we present a deterministic algorithm that explores any tree with $n$ nodes and depth $D$ in $O(\OPT+E)$ rounds, where $\OPT = \Theta(n/k+D)$ is the offline benchmark, and we show that there is no deterministic algorithm that admits an $O(\OPT)+o(E)$ guarantee. For the multiplicative distortion, we present several algorithms that induce a competitive ratio in $\widetilde O(\min\{
\sqrt\mu\log k,
        \mu\log\log k+\mu^{1.5}
    \})$, where the $\widetilde O(\cdot)$ may hide logarithmic factors in $\mu$.
\end{abstract}

\pagenumbering{gobble}
\newpage
\pagenumbering{arabic}

\section{Introduction}
\label{sec:introduction}
Parallel exploration is one of the most fundamental tasks in robotics and artificial intelligence (AI). For example, in robotics, a group of robots entering a building, tunnel system, or underground network must discover the environment while deciding, in real time, how to divide the unexplored regions among themselves. In AI, a group of autonomous agents working on a complex task may need to explore a large combinatorial space, e.g., by collectively probing different combinations of actions.
In both settings, parallelism can dramatically reduce the time required for exploration, but only if the agents' efforts do not overlap excessively.

One mathematical model of this challenge is the \emph{collective tree exploration} problem (CTE) ~\citep{fraigniaud2006collective}. In this problem, a team of $k$ mobile agents starts at the root of an unknown, unweighted tree $T$, with the goal of visiting every vertex as quickly as possible. Time proceeds in synchronous rounds. In each round, every robot may traverse one edge or remain in place. In the centralized full-communication setting that we consider, robots can communicate and compute at no cost. Thus, when a robot visits a vertex, the neighbors of the vertex are revealed to the team. Note that tree exploration in fact captures the more general problem of exploring every vertex in a graph, as one could always explore a spanning tree of the graph.\footnote{This holds naturally if we assume that the vertices are labeled in some canonical way so that a robot can recognize a backward edge.}

Collective tree exploration has typically been studied within the classical worst-case framework of competitive analysis, in which the performance of an online algorithm is compared with the optimal offline completion time $\OPT$, which corresponds to the minimum number of rounds required by $k$ agents to visit all nodes of the tree, given a map in advance. Although computing $\OPT$ exactly is NP-hard~\cite{fraigniaud2006collective}, it holds that $\OPT=\Theta\left(\frac{n}{k}+D\right)$, where $n$ is the number of vertices in the tree and $D$ is its depth.\footnote{In parallel scheduling terminology, the first term corresponds to the total work (serial execution time) divided by the number of workers (processors), while the second term is the length of the longest dependency chain (span). Similar bounds appear in the literature on work stealing; see, e.g., \cite{blumofe1999scheduling}.\label{fn:detail}} The best algorithm to date is $O(\log^2 k)$-competitive~\cite{DBLP:journals/corr/abs-2609-26789}.

In real-world applications, exploration tasks often come with side information, which may be inaccurate. This type of information is already used heuristically in empirical work. For example, frontier-based methods in robotics balance the cost of reaching a point of the frontier against an estimate of the information available beyond that point, and multi-robot systems use such 
information 
to reduce duplicated work~\citep{yamauchi1997frontier,burgard2005coordinated}. More global guidance may come from a floor plan, a building outline, a hand-drawn sketch, a coarse topo-metric graph, or a map from an earlier deployment~\citep{osswald2016speeding,strom2015predictive,luperto2020robot}. In some scenarios, learning systems also predict the unseen portions of occupancy maps and estimate the information gain associated with candidate viewpoints~\citep{shrestha2019learned,SaroyaEtAl2020TunnelPredictions,tao2023seer,TanEtAl2026FourCNet,LupertoEtAl2026MapCompleteness}. 

Despite the widespread use of such side information for exploration problems, to the best of our knowledge, no theoretical analysis quantifies its benefits, particularly when the information is inaccurate. We study this question through the lens of \emph{algorithms with predictions}, also known as \emph{learning-augmented algorithms}~\citep{mitzenmacher2021algorithms}. In this framework, an algorithm receives auxiliary information, referred to as the ``predictions,'' and the goal is to design algorithms that perform well when the predictions are accurate while degrading gracefully with inaccuracy. Formally, our model is as follows.

\paragraph{Subtree-size predictions and error measures.} In this paper, we consider predictions that are estimates of the size of each subtree. 
For a vertex $u$, let $T_u$ be the subtree rooted at $u$, let $V(T_u)$ denote the set of vertices in $T_u$, and let $n_u=|V(T_u)|$ be the size of $T_u$. When the parent of $u$ is visited for the first time by one of the $k$ robots, the algorithm learns about the existence of node $u$ and also 
receives a nonnegative prediction $\hat n_u$ of $n_u$. Without loss of generality, we assume that predictions are all positive integers (see  \cref{lem:integral-predictions} for details).

Different applications make different types of prediction error plausible, and no single scalar measure captures all of their algorithmic effects. We focus primarily on two complementary measures. The \emph{maximum additive error}
\[E:=\max_{u\in V(T)}|\hat n_u-n_u|\]
measures the largest absolute error among the predictions, and the \emph{multiplicative distortion}
\[
\mu := \left(\max_{u\in V(T)}\frac{\hat n_u}{n_u}\right) \cdot \left(\max_{u\in V(T)}\frac{n_u}{\hat n_u}\right)
\]
measures the maximum relative error among the predictions.

Subtree-size predictions are particularly natural given the analogy with parallel scheduling (see footnote \ref{fn:detail} on page \pageref{fn:detail}), which lets us interpret $n_u$ as the total work required to process all nodes in $T_u$. We observe that action predictions, which are common in learning-augmented algorithms \cite{antoniadis2023online}, do not seem as natural.

\subsection{Our Results and Techniques}
Following the learning-augmented viewpoint, our goal is to use accurate advice to improve upon the prediction-free guarantee, ensure that performance degrades smoothly with a problem-relevant error measure, and retain a worst-case guarantee when the advice is arbitrary~\citep{lykouris2018competitive,purohit2018improving,mitzenmacher2021algorithms}. In the terminology of this literature, these desiderata are called \emph{consistency}, \emph{smoothness}, and \emph{robustness}. Consistency measures performance under perfectly accurate predictions, smoothness describes how performance degrades as a function of the error, and robustness captures performance when the error becomes arbitrarily large.

Up to constant factors, the consistency--robustness tradeoff of collective tree exploration is straightforward. For consistency, exact predictions make it possible to achieve $O(n/k+D)$ rounds by partitioning an Eulerian tour of the tree into $k$ segments of length $\Theta(n/k)$ and assigning one segment to each robot (see \cref{lem:perfect-prediction}). For robustness, the problem admits a so-called \emph{min operator}: if two algorithms $A$ and $B$ have completion times $t_A$ and $t_B$, respectively, then there is an algorithm that completes the exploration in at most $4\min\{t_A,t_B\}$ rounds (see \cref{lem:min}). Consequently, any prediction-augmented algorithm can be combined with a prediction-free algorithm to obtain robustness. 
Accordingly, we focus primarily on how performance degrades as a function of the prediction error.\footnote{See Appendix~\ref{appendix:consistency-robustness}
for a brief discussion of why the Eulerian tour algorithm achieving optimal consistency may become ill defined under imperfect predictions, while natural
repairs can lead to poor dependence on the prediction error.} For convenience, we omit the obvious minimization over prediction-free algorithms in our stated bounds. 

Our first result concerns predictions under the additive error model. We present an algorithm whose cost increases linearly with the maximum additive error $E$ and show that a sublinear dependence on $E$ is impossible. Admittedly, the scope of our lower bound is somewhat limited, as it does not rule out, for example, algorithms with a guarantee of the form $O(\OPT + E/k)$ or $\OPT + O(D+E)$. This is further discussed in the conclusion.  

\begin{restatable}[Maximum Additive Error]{theorem}{maximumadditiveerror}
\label{thm:maximum}
There is an algorithm with subtree-size predictions that completes the exploration in $O(\OPT+E)$ rounds when the predictions have (unknown) additive error $E$. 
Furthermore, no deterministic algorithm
admits a uniform completion-time guarantee of
$O(\OPT)+o(E)$ rounds.
\end{restatable}

The algorithm is based on a simple \emph{seed-and-sweep} idea that is developed in Section \ref{sec:maximum}.
In a first \emph{seeding} phase, the
subtree-size predictions are used to disperse the robots so that every
large region of the tree contains its fair share of ``seeds''. In the second \emph{sweep} phase, every robot simply performs a depth-first
exploration starting from its seed, eventually backtracking toward the root
and picking up unexplored branches along the way.  

A key point of the analysis of the seeding phase is that imperfect advice should not be used naively to allocate robots proportionally across branches. For example, when $E=n/k$, a subtree containing essentially no work may still be overestimated by $n/k$, and hence deserve one of the $k$ seeds. The seeding procedure avoids such spurious allocations by first turning the predictions into one-sided, \textit{optimistic estimates} of the work that is guaranteed to be present. It then translates these fractional estimates into integral numbers of robots using a carefully chosen \emph{biased rounding} rule. The crucial property of this rounding is that the error incurred at one branching point does not compound as the robots move deeper into the tree. As a result, prediction errors lead only to a constant loss in the number of seeds assigned to any subtree, rather than an error growing with depth.

Our second result, stated below, concerns predictions under the multiplicative error model.

\begin{theorem}[Multiplicative Distortion]
\label{thm:multiplicative}
There is an algorithm with subtree-size predictions that has a competitive ratio of
    \[ 
        \widetilde O(\min (\mu \log \log k + \mu^{1.5}, \sqrt \mu \log k))
    \] when the predictions have (unknown) multiplicative distortion $\mu$, and where $\tilde O(\cdot)$ hides $\poly \log (1+\mu)$ factors.
\end{theorem}

The main ingredient of our multiplicative distortion algorithms is the consecutive exploration of increasing \emph{skeletons}. For a threshold $\tau$, the $\tau$-skeleton $\mc S_\tau$ is the largest connected component containing the root in which all vertices have predictions larger than $\tau$ (see \cref{fig:threshold-skeleton-components} in \cref{sec:multiplicative} for an illustration). 

The skeletons have two important properties. First, once we have revealed $\mc S_\tau$ for $\tau$ sufficiently small, we can efficiently complete the exploration of the entire tree, for example, by treating trees hanging from the skeleton as complete jobs attached to the skeleton's Eulerian tour and splitting the work as evenly as possible. Second, skeletons can be expanded iteratively and efficiently. For example, suppose we have $\mc S_\tau$ for some $\tau \ge 2n/k$. Then $\mc S_{\tau/2}$ can be determined in $O(\mu D)$ rounds, since one can show that each connected subtree in $\mc S_{\tau/2} \setminus \mc S_\tau$ has at most $O(\mu D)$ vertices. 

This sketched analysis would entail an $O(\mu \log k)$-competitive algorithm; however, we accelerate the expansion of skeletons by (i) applying existing algorithms that do not use predictions to subtrees between skeletons and (ii) expanding the skeleton much more aggressively (not just by a factor of $2$) at early stages to achieve our improved final bounds.

\subsection{Related Work}
\label{sec:related-work}

Collective tree exploration was introduced by \cite{fraigniaud2006collective} along with two communication models:
the complete communication model (the main focus of this paper), and a distributed model in which robots may communicate only by reading and writing information on whiteboards present at all nodes. Following the framework of competitive analysis, they proposed to compare the time required by an online algorithm to explore an unknown tree to the time required by an offline algorithm to visit all nodes of the tree. The latter, denoted $\OPT$, satisfies $\OPT = \Theta(\frac{n}{k}+D)$. Up to a constant factor, the competitive ratio $c(k)$ of a collective exploration algorithm is thus traditionally defined as 
\[
c(k) = \sup_{T} \left\{\frac{\ALG(T,k)}{n/k+D} \right\}
\]
where the supremum is taken over all trees (of arbitrary depth and size). \cite{fraigniaud2006collective} obtained an 
$O(k/\log k)$-competitive algorithm implementable in both communication models and a general constant lower bound on the competitive ratio of any collective tree algorithm. 
The lower bound was then improved to $\Omega(\log k / \log \log k)$ by \cite{dynia2007why,disser2020general}, and the competitive ratio in the complete-communication model was later improved to $O(\sqrt{k})$ \citep{cosson2024treemining,cosson2024potential}, and, very recently, to $O(\log^2 k)$ \cite{DBLP:journals/corr/abs-2609-26789}.

Collective tree exploration has also been investigated in various other regimes, such as under additional constraints \cite{dynia2006power}, for restricted classes of trees \cite{dynia2006smart}, with nonlinear strategies \cite{ortolf2014recursive}, and in the many-agent regime \cite{dereniowski2015fast}. Other objectives than the competitive ratio have also been considered. In particular, \cite{brass2011multirobot,cosson2024treemining,cosson2024potential} primarily aim to design low-regret algorithms, i.e., algorithms with a guarantee of the form $\OPT + f(k,D)$, where $f(k,D)$ is called the regret or the additive overhead.

Of particular relevance to our work is \cite{dereniowski2015fast}, which considers the many-agent regime $k\geq Dn^{1+\epsilon}$ and gives an algorithm that explores the tree in $O(D)$ time steps. Their algorithm sends the robots in successive waves, with each wave collecting increasingly precise information about deeper portions of the tree. At a high level, this information can be interpreted as providing estimates of subtree sizes, which are then used to guide subsequent waves of robots. Although their setting is substantially different from ours---in their regime, the number of robots grows with the size of the tree, whereas we fix $k$ and allow trees of arbitrary size---the rounding argument underlying our additive guarantee is closely related to the rounding scheme used in their analysis.

The modern learning-augmented framework has been studied across a variety of algorithmic domains, including online algorithms, data structures, graph problems, streaming algorithms, mechanism design, and scheduling. Early work developed this framework for problems such as caching, ski rental, scheduling, Bloom filters, and frequency estimation \cite{lykouris2018competitive,purohit2018improving,KraskaBCDP18,Mitzenmacher18,hsu2019learning}; see \cite{mitzenmacher2021algorithms} for an overview. By now, learning-augmented algorithms form an active area of research, and we refer the reader to the Algorithms with Predictions repository for a broad collection of recent and ongoing work in this area.\footnote{\url{https://algorithms-with-predictions.github.io/}} 

One distinguishing feature of our model is that the predictions are not revealed to the algorithm at the start of the execution, but are instead revealed gradually, in an order that depends on the algorithm's past actions. A related line of work is robust flow-time scheduling \cite{azar2021flow,azar2022distortion,gupta2026optimal}, in which multiplicative estimates of the jobs' processing times are revealed at their start date. 

For online graph problems with predictions, \cite{bernardini2022universal} proposed a general error measure that covers online list and routing problems. \cite{eberle2022robustification} studies learning-augmented online graph exploration with a single searcher in an edge-weighted graph and uses
a predicted tour or spanning tree to suggest the next vertex. \cite{banerjee2023searching,depavia2024learning,cabello2024searching} study graph search with predictions, primarily on trees, where a single agent seeks a hidden target and each visited vertex provides a prediction of its distance to the target. In contrast, in our setting, the advice estimates the amount of unexplored work in the subtree rooted at each newly revealed vertex, and the search is carried out by $k$ robots. Thus, the algorithm must use the predictions not only to decide where to explore next, but also to allocate and dynamically reallocate the robots across multiple competing frontiers.

\section{Preliminaries}
\label{sec:preliminaries}
Let $T=(V,E)$ be a finite rooted, unweighted tree with root $r$, $n=|V|$
vertices, and depth $D$. For a vertex $u$, let $T_u$ be the subtree rooted at
$u$, let $n_u:=|V(T_u)|$, let $\operatorname{ch}(u)$ denote its set of children, and let $p(u)$ denote its parent.
There are $k\geq1$ robots, all initially located at $r$.  Time is divided into
synchronous rounds.  In each round, every robot may traverse one incident edge
or remain in place.  When a vertex is first visited, its children and their
incident edges are revealed to every robot.  Communication and computation
are free, so the robots share the complete explored subtree and its frontier
at all times. An exploration is complete when every edge has been traversed.

When a child $v$ is revealed, the algorithm also receives a
prediction $\hat n_v \in \mathbb R_{\ge 0}$ for $n_v$; the root prediction is available at
time zero. No consistency among predictions at different vertices is
assumed, i.e., it is allowed that $\hat n_u \ne \sum_{v \in \ch(u)} \hat n_v+1$. The additive error $E$ and multiplicative distortion $\mu$ are defined as follows:
\begin{align*}
    E&:=\max_{u\in V}|\hat n_u-n_u|,\\
    \mu&:= \left(\max_{u\in V(T)}\frac{\hat n_u}{n_u}\right) \cdot \left(\max_{u\in V(T)}\frac{n_u}{\hat n_u}\right),
\end{align*}
where division by $0$ is interpreted as infinity. We denote by $\mu_e = \max_{u\in V(T)}\frac{\hat n_u}{n_u}$ the expansion and $\mu_c = \max_{u\in V(T)}\frac{n_u}{\hat n_u}$ the contraction, so $\mu = \mu_e \mu_c$. 

We make the following assumptions: $\mu_e$ and $\mu_c$ are at least $1$; the predictions are positive integers $\hat n_u \in \Z_+$ for all $u$; and the predictions are monotone, i.e., $\hat n_u \le \hat n_{p(u)}$ for all $u \ne r$. We show in \cref{appendix:assumption} that these assumptions are without loss of generality.

We let $\OPT(T,k) = \Theta(n/k + D)$ denote the minimum completion time of an offline algorithm
that knows $T$ in advance. We write it simply as $\OPT$ when the instance is
clear.

\section{The Maximum Additive Error}
\label{sec:maximum}
In this section, we consider predictions with a bounded additive error and prove \Cref{thm:maximum}, which we restate below. 
\maximumadditiveerror*
Our algorithm, called \textsc{Seed-And-Sweep} (\Cref{alg:sas}), operates in two phases. In the first phase, each robot reaches a `seed', which is a node defined using only the advice available at visited nodes and reached without the robot ever backtracking. In the second phase, the robot performs a depth-first search that starts from its seed and backtracks along edges previously discovered by other robots. The algorithm is illustrated in \Cref{fig:seed-sweep}. 

To analyze the algorithm, we show that when the list of seeds $\{s_i\}_{i\in[k]}$ is sufficiently well-dispersed in the tree $T$, the algorithm \textsc{Seed-And-Sweep} has fast exploration guarantees. This is formalized in Lemma~\ref{lem:scatter} below (\textit{dispersed seeds imply a fast sweep}). Then, we show that, if the advice has bounded additive error, the robots can efficiently disperse in the tree. This is formalized in Lemma \ref{lem:xhat} (\textit{good advice induces dispersed seeds}). 

Specifically, Lemma \ref{lem:xhat} relies on an explicit seeding procedure called \textsc{Biased Rounding Seeding} (Algorithm~\ref{alg:sands}), in which the robots partition themselves roughly in proportion to the estimated sizes of the dangling subtrees. For this procedure, we assume that the prediction error is one-sided, i.e., $n_u-E \leq \hat n_u \leq n_u$, which is without loss of generality (see Remark \ref{rem:doubling-trick}). For an integer $k\in \Nbb$ and a ratio $\alpha \in [0,1]$, we define the \textit{biased rounding}
\begin{equation}
    f(\alpha, k) = \begin{cases}
    \min\{\ceil{\alpha k}+1,k\} \quad \text{if}\quad 1/2<\alpha \leq 1\\
    \max\{\floor{\alpha k}-1,0\} \quad \text{if}\quad \alpha \leq 1/2.
\end{cases}
\end{equation}
Observe that $f(\cdot,\cdot)$ satisfies $\sum_u f(\alpha_u, k)\leq k$ whenever $\sum_u \alpha_u \leq 1$.

\begin{algorithm}[H]
\caption{\textsc{Seed-And-Sweep}}
\label{alg:sas}
\begin{algorithmic}[1]
\Statex \hspace{-\algorithmicindent}\textbf{Execution:} The algorithm is executed in parallel for all robots $i\in[k]$
\State \textbf{Go to} the starting seed $s_i$ \Comment{Phase 1: Seed (see Algorithm \ref{alg:sands})}
\While{adjacent to an unexplored child or not located at the root} \Comment{Phase 2: Sweep}
\If{adjacent to an unexplored child}
\State Move to that child
\Else
\State Move to the parent of the current node
\EndIf
\EndWhile
\end{algorithmic}
\end{algorithm}

\begingroup


\definecolor{sweepRed}{RGB}{230,159,0}      
\definecolor{sweepBlue}{RGB}{0,114,178}     
\definecolor{sweepGreen}{RGB}{204,121,167}  


\tikzset{treefig/.style={
  xscale=1.25,
  line cap=round,
  line join=round,
  every node/.style={font=\normalfont\small}
}}

\tikzset{panelbox/.style={
  draw=gray!38,
  fill=gray!3,
  rounded corners=8pt,
  inner sep=9pt
}}

\tikzset{edge/.style={
  draw=gray!72!black,
  line width=1.25pt
}}

\tikzset{rededge/.style={
  draw=sweepRed!86!black,
  line width=1.45pt
}}

\tikzset{blueedge/.style={
  draw=sweepBlue!88!black,
  line width=1.45pt,
  dash pattern=on 4.5pt off 2.2pt
}}

\tikzset{greenedge/.style={
  draw=sweepGreen!82!black,
  line width=1.45pt,
  dash pattern=on 5pt off 1.8pt on 1.2pt off 1.8pt
}}

\tikzset{treevertex/.style={
  circle,
  line width=1.05pt,
  inner sep=0pt,
  minimum size=4.8mm
}}

\tikzset{filled/.style={
  treevertex,
  draw=gray!72!black,
  fill=white
}}

\tikzset{hollow/.style={
  treevertex,
  draw=gray!72!black,
  fill=white
}}

\tikzset{rednode/.style={
  treevertex,
  draw=sweepRed!86!black,
  fill=sweepRed!18
}}

\tikzset{bluenode/.style={
  treevertex,
  draw=sweepBlue!88!black,
  fill=sweepBlue!12
}}

\tikzset{greennode/.style={
  treevertex,
  draw=sweepGreen!82!black,
  fill=sweepGreen!14
}}

\tikzset{guidearrow/.style={
  -{Latex[length=2.7mm,width=2.0mm]},
  shorten <=2.7mm,
  shorten >=2.7mm
}}

\tikzset{treelabel/.style={
  fill=white,
  inner sep=1.2pt
}}


\newcommand{\treecoordinates}{%
\coordinate (root) at (0.00,6.50);%

\coordinate (a) at (-0.65,5.25);%
\coordinate (p3) at (0.65,5.25);%

\coordinate (b) at (-1.20,4.00);%
\coordinate (p2) at (-0.10,4.00);%
\coordinate (p3a) at (1.10,4.00);%

\coordinate (p1) at (-1.75,2.75);%
\coordinate (bleaf) at (-0.65,2.75);%
\coordinate (p2j) at (0.10,2.75);%
\coordinate (p3b) at (1.40,2.75);%

\coordinate (p1a) at (-2.20,1.50);%
\coordinate (p1b) at (-1.30,1.50);%
\coordinate (p2l) at (-0.35,1.50);%
\coordinate (p2r) at (0.55,1.50);%
\coordinate (p3c) at (1.60,1.50);%

\coordinate (p1aa) at (-2.35,0.25);%
\coordinate (p1ba) at (-1.65,0.25);%
\coordinate (p1bb) at (-0.95,0.25);%
}


\newcommand{\seedtreeedges}{%

\draw[edge]
    (root)
    .. controls (-0.15,6.10) and (-0.45,5.65) ..
    (a);%

\draw[edge]
    (a)
    .. controls (-0.82,4.85) and (-1.02,4.40) ..
    (b);%

\draw[edge]
    (b)
    .. controls (-1.40,3.65) and (-1.60,3.10) ..
    (p1);%

\draw[edge]
    (b)
    .. controls (-1.00,3.65) and (-0.75,3.10) ..
    (bleaf);%

\draw[edge]
    (p1)
    .. controls (-1.95,2.40) and (-2.12,1.90) ..
    (p1a);%

\draw[edge]
    (p1a)
    .. controls (-2.25,1.10) and (-2.32,0.65) ..
    (p1aa);%

\draw[edge]
    (p1)
    .. controls (-1.55,2.40) and (-1.38,1.90) ..
    (p1b);%

\draw[edge]
    (p1b)
    .. controls (-1.40,1.10) and (-1.58,0.65) ..
    (p1ba);%

\draw[edge]
    (p1b)
    .. controls (-1.15,1.10) and (-1.02,0.65) ..
    (p1bb);%

\draw[edge]
    (a)
    .. controls (-0.50,4.85) and (-0.25,4.40) ..
    (p2);%

\draw[edge]
    (p2)
    .. controls (-0.05,3.60) and (0.05,3.15) ..
    (p2j);%

\draw[edge]
    (p2j)
    .. controls (-0.05,2.40) and (-0.25,1.90) ..
    (p2l);%

\draw[edge]
    (p2j)
    .. controls (0.25,2.40) and (0.45,1.90) ..
    (p2r);%

\draw[edge]
    (root)
    .. controls (0.15,6.10) and (0.45,5.65) ..
    (p3);%

\draw[edge]
    (p3)
    .. controls (0.80,4.82) and (1.00,4.42) ..
    (p3a);%

\draw[edge]
    (p3a)
    .. controls (1.20,3.62) and (1.32,3.15) ..
    (p3b);%

\draw[edge]
    (p3b)
    .. controls (1.48,2.35) and (1.56,1.92) ..
    (p3c);%
}


\newcommand{\sweeptreeedges}{%

\draw[blueedge]
    (root)
    .. controls (-0.15,6.10) and (-0.45,5.65) ..
    (a);%

\draw[blueedge]
    (a)
    .. controls (-0.82,4.85) and (-1.02,4.40) ..
    (b);%

\draw[rededge]
    (b)
    .. controls (-1.40,3.65) and (-1.60,3.10) ..
    (p1);%

\draw[blueedge]
    (b)
    .. controls (-1.00,3.65) and (-0.75,3.10) ..
    (bleaf);%

\draw[rededge]
    (p1)
    .. controls (-1.95,2.40) and (-2.12,1.90) ..
    (p1a);%

\draw[rededge]
    (p1a)
    .. controls (-2.25,1.10) and (-2.32,0.65) ..
    (p1aa);%

\draw[rededge]
    (p1)
    .. controls (-1.55,2.40) and (-1.38,1.90) ..
    (p1b);%

\draw[rededge]
    (p1b)
    .. controls (-1.40,1.10) and (-1.58,0.65) ..
    (p1ba);%

\draw[rededge]
    (p1b)
    .. controls (-1.15,1.10) and (-1.02,0.65) ..
    (p1bb);%

\draw[blueedge]
    (a)
    .. controls (-0.50,4.85) and (-0.25,4.40) ..
    (p2);%

\draw[blueedge]
    (p2)
    .. controls (-0.05,3.60) and (0.05,3.15) ..
    (p2j);%

\draw[blueedge]
    (p2j)
    .. controls (-0.05,2.40) and (-0.25,1.90) ..
    (p2l);%

\draw[blueedge]
    (p2j)
    .. controls (0.25,2.40) and (0.45,1.90) ..
    (p2r);%

\draw[greenedge]
    (root)
    .. controls (0.15,6.10) and (0.45,5.65) ..
    (p3);%

\draw[greenedge]
    (p3)
    .. controls (0.80,4.82) and (1.00,4.42) ..
    (p3a);%

\draw[greenedge]
    (p3a)
    .. controls (1.20,3.62) and (1.32,3.15) ..
    (p3b);%

\draw[greenedge]
    (p3b)
    .. controls (1.48,2.35) and (1.56,1.92) ..
    (p3c);%
}


\newcommand{\seednodes}{%

\node[filled] at (root) {};%

\node[filled] at (a) {};%
\node[filled] at (b) {};%

\node[rednode] at (p1) {};%
\node[hollow] at (p1a) {};%
\node[filled] at (p1aa) {};%

\node[hollow] at (p1b) {};%
\node[hollow] at (p1ba) {};%
\node[hollow] at (p1bb) {};%

\node[hollow] at (bleaf) {};%

\node[bluenode] at (p2) {};%
\node[filled] at (p2j) {};%
\node[hollow] at (p2l) {};%
\node[hollow] at (p2r) {};%

\node[greennode] at (p3) {};%
\node[hollow] at (p3a) {};%
\node[filled] at (p3b) {};%
\node[filled] at (p3c) {};%
}


\newcommand{\sweepnodes}{%

\node[greennode] at (root) {};%

\node[bluenode] at (a) {};%
\node[bluenode] at (b) {};%

\node[rednode] at (p1) {};%
\node[rednode] at (p1a) {};%
\node[rednode] at (p1aa) {};%
\node[rednode] at (p1b) {};%
\node[rednode] at (p1ba) {};%
\node[rednode] at (p1bb) {};%

\node[bluenode] at (bleaf) {};%
\node[bluenode] at (p2) {};%
\node[bluenode] at (p2j) {};%
\node[bluenode] at (p2l) {};%
\node[bluenode] at (p2r) {};%

\node[greennode] at (p3) {};%
\node[greennode] at (p3a) {};%
\node[greennode] at (p3b) {};%
\node[greennode] at (p3c) {};%
}


\newcommand{\treelabels}{%

\node[treelabel]
    at (0.22,6.82) {$r$};%

\node[treelabel]
    at (-2.10,3.02) {$s_1$};%

\node[treelabel]
    at (0.31,4.18) {$s_2$};%

\node[treelabel]
    at (1.02,5.52) {$s_3$};%
}


\newcommand{\seedoverlay}{%

\draw[guidearrow,rededge]
    (root) to[out=225,in=85] (p1);%

\draw[guidearrow,blueedge]
    (root) to[out=255,in=95] (p2);%

\draw[guidearrow,greenedge]
    (root) to[out=325,in=105] (p3);%
}


\newcommand{\treepanelbackground}{%
\begin{scope}[on background layer]
  \node[
    panelbox,
    fit=(root)(p1aa)(p1bb)(p3c)
  ] {};
\end{scope}%
}


\newcommand{\seedpicture}{%
\begin{tikzpicture}[treefig]%
\treecoordinates%
\seedtreeedges%
\seednodes%
\seedoverlay%
\treelabels%
\end{tikzpicture}%
}

\newcommand{\sweeppicture}{%
\begin{tikzpicture}[treefig]%
\treecoordinates%
\sweeptreeedges%
\sweepnodes%
\treelabels%
\end{tikzpicture}%
}


\begin{figure}[h!]
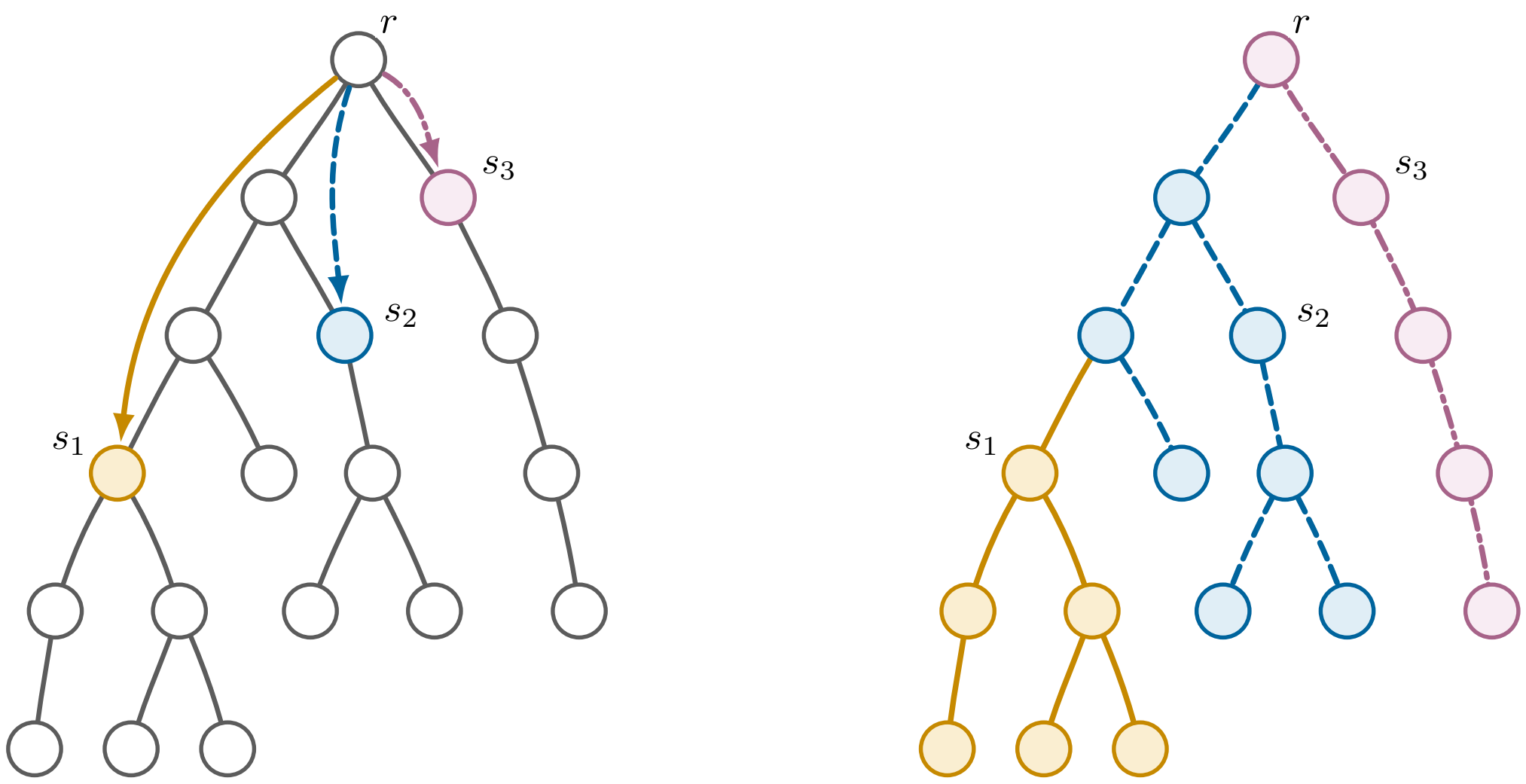

\centering

\makebox[\textwidth][c]{%
\begin{subfigure}[t]{0.32\textwidth}
\centering
\resizebox{\linewidth}{!}{%
    \seedpicture%
}
\caption{Seed}
\label{fig:seed}
\end{subfigure}
\hspace{0.15\textwidth}
\begin{subfigure}[t]{0.32\textwidth}
\centering
\resizebox{\linewidth}{!}{%
    \sweeppicture%
}
\caption{Sweep}
\label{fig:sweep}
\end{subfigure}
}

\caption{Seed and sweep. In the sweep panel, each edge is styled
according to the first robot that traverses it: solid orange for $s_1$,
dashed blue for $s_2$, and dash-dotted purple for $s_3$.}
\label{fig:seed-sweep}

\end{figure}

\endgroup

\begin{lemma}[Dispersed seeds imply a fast sweep]\label{lem:scatter}
Consider \Cref{alg:sas} (\textsc{Seed-And-Sweep}) and denote by $k_u$ the number of robots whose seed is in $T(u)$. If there are $c_1\in(0,1]$ and $c_2>0$ satisfying 
\begin{equation}\label{eq:lb-seed}
    \forall u\in T: k_u \geq c_1 k\frac{n_u}{n}- c_2,
\end{equation}
then all nodes are explored by time
\[
\left(\frac{1+2c_2}{c_1}\right)\frac{2n}{k}+D.
\]
\end{lemma}
\begin{proof}
Denote by $S = \{s_i\}_{i\in [k]}$ the multiset of all initial seed locations and consider the Steiner tree of $S \cup \{r\}$, which corresponds to the tree consisting of the nodes $u\in T$ such that $k_u\geq 1$. Consider a node $u$ with $k_u\geq 1$ that has a child $v$ with $k_v=0$. 

If we can prove that by time $t_v$ a robot reaches node $v$, then, by the property of depth-first search in the subtree rooted at $v$, that subtree must be explored entirely by time \[t_v+2n_v \leq t_v + \frac{c_2}{c_1} \frac{2n}{k},\] 
where we used $k_v = 0$ in \eqref{eq:lb-seed}. 

Let us now provide a bound on $t_v$, for any such choice of $v$. There are $k_u$ robots that are seeded in the subtree rooted at $u$. Clearly, none of those can return to the parent of $u$ before $v$ has been explored. Furthermore, all movements away from the root correspond to discovering a new node below $u$ that is not below $v$. Thus, after $t$ steps, at least $k_u(t-D)/2$ of the $n_u-n_v$ nodes in $T(u)\setminus T(v)$ have been explored. Consequently, 
\[
t_v\leq \frac{2n_u}{k_u}+D \leq \left(\frac{1+c_2}{c_1}\right)\frac{2n}{k} +D
\]
where we used \eqref{eq:lb-seed} and $k_u\geq 1$. This completes the proof. 
\end{proof}

\begin{algorithm}[H]
\caption{\textsc{Biased Rounding Seeding}}
\label{alg:sands}
\begin{algorithmic}[1]
\Statex \hspace{-\algorithmicindent}\textbf{Input:} Each node $u\in T$ receives an estimate $\hat n_u$ satisfying $n_u-E\leq \hat n_u \leq n_u $
\Statex \hspace{-\algorithmicindent}\textbf{Execution:} The algorithm is executed in parallel on non-leaf nodes $p\in T\setminus \Lcal$
\Statex \hspace{-\algorithmicindent}\textbf{Output:} All robots reach a seed in at most $D$ steps
\While{there is more than one robot at $p\in T\setminus\Lcal$}
\State Define $\hat x_p\geq 2$ to be the number of robots at $p$
\State Define $\hat \alpha_u = \frac{\hat n_u}{\sum_{u\in \ch(p)}\hat n_u}$ for all children $u\in \ch(p)$
\State Send $\hat x_u = f(\hat \alpha_u,\hat x_p)$ robots to $u$, where $f(\alpha, k) = \begin{cases}
    \min\{\ceil{\alpha k}+1,k\} \quad \text{if}\quad 1/2<\alpha \leq 1\\
    \max\{\floor{\alpha k}-1,0\} \quad \text{if}\quad \alpha \leq 1/2.\end{cases}$
\EndWhile \vspace{0.1cm}
\end{algorithmic}
\end{algorithm}

\begin{lemma}[Good advice induces dispersed seeds]\label{lem:xhat}
    If $E\leq \frac{n}{k}$, then there is a constant $c$ such that \Cref{alg:sands} satisfies
    \begin{equation}\label{eq:bound}
        \forall u\in T:\qquad k_u \geq k\frac{n_u}{n} - c.
    \end{equation}
\end{lemma}
\begin{proof}
We prove \eqref{eq:bound} by induction on the nodes of the tree, starting from the root. Equation \eqref{eq:bound} is clearly verified at the root, as $k_r = k$. Assume \eqref{eq:bound} holds at some node $p\in T\setminus \Lcal$ and consider one of its children $u \in \ch(p)$. Our goal is to show that \eqref{eq:bound} holds at $u$.

First note that
\begin{align}
    \hat \alpha_u  \geq \alpha_u - \frac{E}{n_p}, \label{eq:alphahat}
\end{align}
which follows from the definition of $\hat \alpha_u = \frac{\hat n_u}{\sum_{v\in C(p(u))}\hat n_v}$ and $\alpha_u = \frac{n_u}{n_p}$, and the assumption that the advice satisfies $\forall v\in T :  n_v -E\leq \hat n_v \leq n_v$.

Using the induction hypothesis, we write
\begin{align}
    \hat \alpha_u k_p 
    \geq \hat \alpha_u \left(k\frac{n_p}{n}-c\right) = k\frac{n_u}{n}-\hat \alpha_uc+(\hat \alpha_u -\alpha_u) k\frac{n_p}{n} \geq k\frac{n_u}{n} - \hat \alpha_uc -1
\end{align}
where, in the last inequality, we used \eqref{eq:alphahat} and $E\leq \frac{n}{k}$.

We finish by considering two cases. First, suppose that $\hat \alpha_u\leq 1/2$. Using \eqref{eq:alphahat}, we obtain  
\begin{align*}
    k_u = f(\hat \alpha_u,k_p)&\geq \hat \alpha_u k_p -2\geq k\frac{n_u}{n} -c/2 -3\geq k \frac{n_u}{n} - c,
\end{align*}
where we used $c\geq 6$. For $\hat \alpha_u\geq 1/2$, if $k_u=k_p$, then the conclusion is obvious; otherwise, using \eqref{eq:alphahat}, 
\begin{align*}
    k_u = f(\hat \alpha_u,k_p)
    \geq \hat \alpha_u k_p +1
    \geq k\frac{n_u}{n} -c -1+1
    = k \frac{n_u}{n} - c.
\end{align*}
\end{proof}
We complete the section with the proof of Theorem \ref{thm:maximum}.

\begin{proof}[Proof of Theorem \ref{thm:maximum}]
First, assume that $E\leq \frac{n}{k}$. In this case, by Lemma \ref{lem:scatter}, \textsc{Biased Rounding Seeding} satisfies the lemma's condition with $c_1 =1$ and $c_2=c$. Therefore, the algorithm's running time is $O(\OPT)$. 

We now assume that $E\geq \frac{n}{k}$. We introduce $k' = \ceil{\frac{n}{E}}$ and consider the variant of \textsc{Biased Rounding Seeding} that starts with $k'$ robots at the root. Since $k'<k$ and $f(\cdot,\cdot)$ is monotone in its second argument, the value of $k_u$ produced by this initialization is a lower bound on the corresponding value produced by initializing \textsc{Biased Rounding Seeding} with $k$ robots. Consequently, Lemma \ref{lem:xhat} gives
\begin{align*}
    k_u &\geq k'\frac{n_u}{n} - c 
    \geq \frac{n_u}{E} -c
    = \frac{n}{kE} k\frac{n_u}{n} -c.
\end{align*}
Thus, the conditions of Lemma \ref{lem:scatter} are met with $c_1 = \frac{n}{kE}$ and $c_2 =c$, so the algorithm's running time is $O(E+D)$.

Thus, the running time of \textsc{Seed-And-Sweep} is $O(\OPT + E)$. 
\end{proof}

We now turn to the lower bound, which follows from a simple linear-scaling argument that itself relies on the prediction-free 
lower bound of \cite{dynia2007why}. The latter only concerns deterministic algorithms, and so does the statement below. 

\begin{lemma}[Lower bound]
    There is no deterministic collective exploration algorithm with a guarantee of $O(\OPT) + o(E)$.
\end{lemma}
\begin{proof} 
    Suppose that there exists an algorithm $\ALG$ with such a guarantee, i.e., one that explores any tree with $n$ nodes and depth $D$ under predictions of error $E$ in $C(\frac{n}{k}+D) + g(E)$ time steps, for a constant $C$ and a function $g(E)$ satisfying $g(E)/E \rightarrow_{E\rightarrow \infty} 0$. We will obtain a contradiction to the classical $\Omega(\log k / \log \log k)$ lower bound on the competitive ratio of collective tree exploration \cite{dynia2007why}.

    Consider a fixed tree $T$ with $n$ nodes and depth $D$ and with subtree size predictions subject to additive error $E$. For an arbitrarily large dilatation factor $\lambda \in \Nbb$, consider the virtual tree $T'$ in which all edges are subdivided into $\lambda$ sub-edges, and for which the predictions are multiplied by $\lambda$. Note that the parameters of the virtual tree $T'$ all scale linearly: $n' = \lambda n$, $D' = \lambda D$, and $E' = \lambda E$.\footnote{The predictions for the newly added vertices in subdivided edges are naturally defined by incrementing the prediction of the lower endpoint of the original edge.} Now consider the algorithm $\ALG'$ that uses algorithm $\ALG$ on the virtual tree $T'$, while physically exploring the real tree $T$ (concretely, $\ALG$ moves once every $\lambda$ moves of $\ALG'$).

    The time required by that algorithm to explore the initial tree can thus be written as $O(\OPT) + \frac{1}{\lambda}f(\lambda E)$, and, for sufficiently large $\lambda$ (e.g., $\lambda$ such that $f(\lambda \hat n_r)/\lambda<1$), the second term is negligible. Thus, the algorithm runs in $O(\OPT)$, even under $E = n_r$, which contradicts the existing prediction-free lower bound of $\Omega(\log k / \log \log k)$ \cite{dynia2007why}. 
\end{proof}

\begin{remark}[One-sided error and doubling trick]\label{rem:doubling-trick} The algorithm presented above implicitly assumes that predictions have one-sided error, which can be obtained from a pre-processing of the form $\hat n_u \gets \hat n_u- E$.
Since \(E\) is unknown, a constant-factor upper bound on $E$ must be guessed via a standard doubling trick that we now explain.
For guesses \(\hat E=1,2,4,\ldots\), run the algorithm with predictions
\(\hat n_u \gets \max\{1,\hat n_u-\hat E\}\).
Whenever \(\hat E\ge E\), we have \(n_u-2\hat E\le \hat n_u\le n_u\) for every \(u\), and the assumption holds with error $2\hat E = O(E)$. Restart the algorithm with the next guess if it has not finished after \(C\hat E\) rounds.
A sufficiently large constant \(C\) guarantees completion at the first guess with \(\hat E\ge\max\{E,\OPT\}\).
Since the running times of preceding algorithms form a geometric series, the total running time remains \(O(\OPT+E)\).
\end{remark}

\begin{remark}[Distributed algorithm] Interestingly, the algorithm presented in this section can be implemented in a particularly stringent distributed communication model in which agents can only tell whether a node was previously explored and no other communication between agents is permitted (this corresponds to allowing one bit of memory in the whiteboard model of \cite{fraigniaud2006collective}).  Indeed, the \textsc{Biased Rounding Seeding} procedure can be implemented without any communication, and the \textsc{Sweep} phase only requires robots to recognize previously explored edges. Conceptually, this means that, under bounded additive error, predictions can replace communication. Note that, in contrast, \cite{fraigniaud2006collective} proved an $\Omega(k)$ lower bound on the competitive ratio of collective tree exploration algorithms without any communication.  
\end{remark}

\section{Multiplicative Distortion}
\label{sec:multiplicative}
In this section, we prove \cref{thm:multiplicative}. The main idea is to explore a nested sequence of \emph{skeletons}. A skeleton contains the portion of the tree that the predictions deem too large to leave as a single unexplored job, while the components added between two consecutive skeletons form the units that we assign to the robots.

The rest of the section is organized as follows. We first assume that the predictions are normalized to be one-sided, i.e., $n_u\leq\hat n_u\leq\mu n_u$, and that the distortion $\mu$ is known to the algorithm. 

In \cref{subsec:multiplicative-framework}, we present the skeleton framework and a simple $O(\mu \log k)$-competitive algorithm as a warm-up. In \cref{subsec:multiplicative-bootstrap} and \cref{subsec:multiplicative-aggressive}, we give two improved algorithms with $O(\sqrt \mu \log k)$ and $\widetilde O(\mu \log \log k + \mu^{1.5})$ competitive ratios, respectively. Finally, in \cref{subsec:multiplicative-assumption}, we remove the assumption that the distortion is known.

\subsection{The skeleton framework and an $O(\mu\log k)$ baseline}
\label{subsec:multiplicative-framework}
For a threshold $\tau \le \hat n_r$, the \emph{$\tau$-skeleton} $\Scal_\tau$ is the connected component containing all vertices whose prediction is at least $\tau$, i.e. $\Scal_\tau := \{u: \hat n_u \ge \tau\}$. 
Note that $\Scal_\tau$ is connected by the monotonicity assumption on $\hat n$. The leaves of a threshold $\tau$ are defined by $\Lcal_\tau:= \{u \in S_\tau :\ch(u) \cap \Scal_\tau = \varnothing\}$. The \emph{frontier} $\mathcal F_\tau$ of the $\tau$-skeleton consists of the unvisited children of its vertices, i.e. \[\Fcal_\tau = \ch(\mc S_\tau) \setminus \mc S_\tau, \text{ where } \ch(\mc S_\tau) = \bigcup_{u \in \mc S_\tau} \ch(u) .\]
Every frontier vertex is the root of a full hanging subtree. The hanging subtrees rooted at the frontier vertices of any given threshold level are pairwise disjoint, and every frontier prediction is smaller than the current threshold.

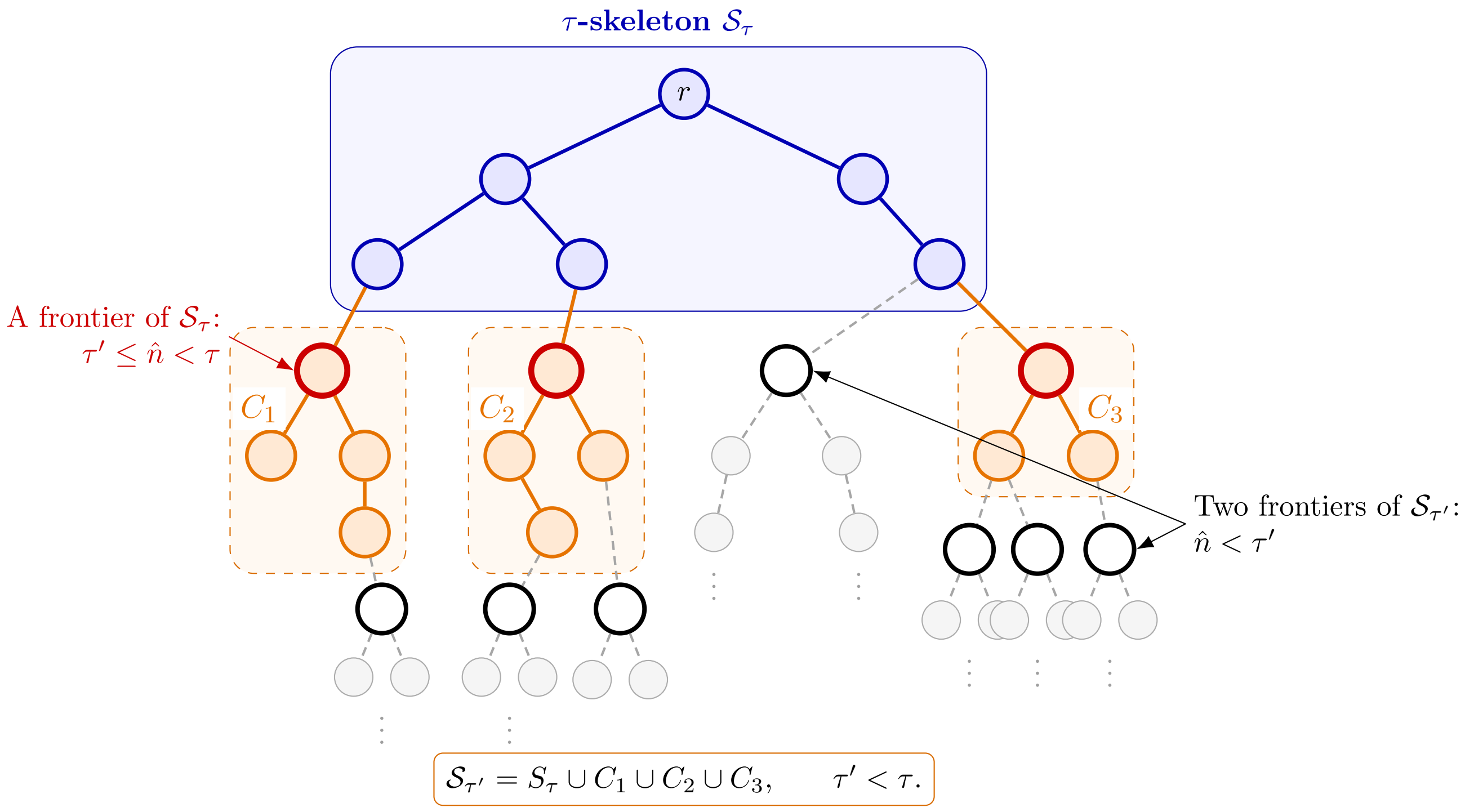
\begin{figure}[t]
  \centering
\begin{tikzpicture}[
  x=1cm,y=1cm,
  every node/.style={font=\small},
  skel/.style={circle,draw=blue!70!black,fill=blue!10,very thick,
               minimum size=5.7mm,inner sep=0pt},
  comp/.style={circle,draw=orange!90!black,fill=orange!17,very thick,
               minimum size=5.7mm,inner sep=0pt},
  oldfront/.style={circle,draw=red!80!black,fill=orange!17,line width=2pt,
                   minimum size=5.9mm,inner sep=0pt},
  frontier/.style={circle,draw=black,fill=white,line width=1.5pt,
                   minimum size=5.7mm,inner sep=0pt},
  hidden/.style={circle,draw=gray!65,fill=gray!8,
                 minimum size=4.5mm,inner sep=0pt},
  skedge/.style={blue!70!black,very thick},
  compedge/.style={orange!90!black,very thick},
  hiddenedge/.style={gray!70,densely dashed,thick},
  note/.style={align=center,fill=white,inner sep=2.5pt},
  >=Latex
]

\node[skel] (r)  at (0,0) {$r$};
\node[skel] (a)  at (-2.1,-1.0) {};
\node[skel] (b)  at ( 2.1,-1.0) {};
\node[skel] (a1) at (-3.6,-2.0) {};
\node[skel] (a2) at (-1.2,-2.0) {};
\node[skel] (b1) at ( 3,-2.0) {};

\draw[skedge] (r)--(a) (r)--(b)
              (a)--(a1) (a)--(a2)
              (b)--(b1) ;

\node[oldfront] (x1) at (-4.25,-3.25) {};
\node[oldfront] (x2) at (-1.50,-3.25) {};
\node[frontier] (x3) at (1.20,-3.25) {};
\node[oldfront] (x4) at (4.25,-3.25) {};

\draw[compedge]   (a1)--(x1) (a2)--(x2) (b1)--(x4);
\draw[hiddenedge] (b1)--(x3);

\node[comp] (c11) at (-4.85,-4.25) {};
\node[comp] (c12) at (-3.75,-4.25) {};
\node[comp] (c13) at (-3.75,-5.15) {};
\draw[compedge] (x1)--(c11) (x1)--(c12) (c12)--(c13);
\node[frontier] (y12) at (-3.55,-6.05) {};
\draw[hiddenedge] (c13)--(y12);

\node[comp] (c21) at (-2.05,-4.25) {};
\node[comp] (c22) at (-0.95,-4.25) {};
\node[comp] (c23) at (-1.55,-5.15) {};
\draw[compedge] (x2)--(c21) (x2)--(c22) (c21)--(c23);
\node[frontier] (y21) at (-2.05,-6.05) {};
\node[frontier] (y22) at (-0.75,-6.05) {};
\draw[hiddenedge] (c23)--(y21) (c22)--(y22);

\node[hidden] (h31) at (0.55,-4.25) {};
\node[hidden] (h32) at (1.85,-4.25) {};
\node[hidden] (h33) at (0.35,-5.15) {};
\node[hidden] (h34) at (2.05,-5.15) {};
\draw[hiddenedge] (x3)--(h31) (x3)--(h32)
                  (h31)--(h33) (h32)--(h34);
\node[text=gray!75] at (0.35,-5.68) {$\vdots$};
\node[text=gray!75] at (2.05,-5.68) {$\vdots$};

\node[comp] (c41) at (3.70,-4.25) {};
\node[comp] (c42) at (4.80,-4.25) {};
\draw[compedge] (x4)--(c41) (x4)--(c42);
\node[frontier] (y41) at (3.35,-5.35) {};
\node[frontier] (y42) at (4.15,-5.35) {};
\node[frontier] (y43) at (5.00,-5.35) {};
\draw[hiddenedge] (c41)--(y41) (c41)--(y42) (c42)--(y43);


\node[hidden] (z121) at (-3.88,-6.85) {};
\node[hidden] (z122) at (-3.22,-6.85) {};
\draw[hiddenedge] (y12)--(z121) (y12)--(z122);
\node[text=gray!75] at (-3.55,-7.37) {$\vdots$};

\node[hidden] (z211) at (-2.38,-6.85) {};
\node[hidden] (z212) at (-1.72,-6.85) {};
\draw[hiddenedge] (y21)--(z211) (y21)--(z212);
\node[text=gray!75] at (-2.05,-7.37) {$\vdots$};

\node[hidden] (z221) at (-1.08,-6.88) {};
\node[hidden] (z222) at (-0.42,-6.88) {};
\draw[hiddenedge] (y22)--(z221) (y22)--(z222);

\node[hidden] (z411) at (3.02,-6.18) {};
\node[hidden] (z412) at (3.68,-6.18) {};
\draw[hiddenedge] (y41)--(z411) (y41)--(z412);
\node[text=gray!75] at (3.35,-6.70) {$\vdots$};

\node[hidden] (z421) at (3.82,-6.18) {};
\node[hidden] (z422) at (4.48,-6.18) {};
\draw[hiddenedge] (y42)--(z421) (y42)--(z422);
\node[text=gray!75] at (4.15,-6.70) {$\vdots$};

\node[hidden] (z431) at (4.67,-6.18) {};
\node[hidden] (z432) at (5.33,-6.18) {};
\draw[hiddenedge] (y43)--(z431) (y43)--(z432);
\node[text=gray!75] at (5.00,-6.70) {$\vdots$};

\begin{scope}[on background layer]
  \node[fit=(r)(a)(b)(a1)(a2)(b1),
        draw=blue!65!black,fill=blue!4,rounded corners=9pt,
        inner sep=7pt,label={[blue!70!black]above:\textbf{$\tau$-skeleton $\mc S_\tau$}}] {};

  \node[fit=(x1)(c11)(c12)(c13),
        draw=orange!85!black,fill=orange!5,dashed,rounded corners=8pt,
        inner sep=5pt] (box1) {};
  \node[fit=(x2)(c21)(c22)(c23),
        draw=orange!85!black,fill=orange!5,dashed,rounded corners=8pt,
        inner sep=5pt] (box2) {};
  \node[fit=(x4)(c41)(c42),
        draw=orange!85!black,fill=orange!5,dashed,rounded corners=8pt,
        inner sep=5pt] (box3) {};
\end{scope}

\node[note,text=orange!90!black,font=\bfseries] at (-4.99,-3.70) {$C_1$};
\node[note,text=orange!90!black,font=\bfseries] at (-2.19,-3.70) {$C_2$};
\node[note,text=orange!90!black,font=\bfseries] at ( 4.95,-3.70) {$C_3$};

\node[note,text=red!80!black,align=right] (oldfrontlabel) at (-6.7,-2.85)
  {A frontier of $\mc S_\tau$:\\$\tau'\le \hat n<\tau$};
\draw[red!80!black,-{Latex[length=2mm]}] (oldfrontlabel.east) -- (x1.west);

\node[note,text=black,align=left] (newfront) at (7.55,-5.05)
  {Two frontiers of $\mc S_{\tau'}$:\\$\hat n <\tau'$};
\draw[black,-{Latex[length=2mm]}] (newfront.west) -- (y43.east);
\draw[black,-{Latex[length=2mm]}] (newfront.west) -- (x3.east);


\node[draw=orange!85!black,fill=white,rounded corners=3pt,inner sep=4pt,
      align=center] at (0,-8.05)
  {$\mc S_{\tau'}=S_\tau\cup C_1\cup C_2\cup C_3$,\qquad $\tau'<\tau$.};

\end{tikzpicture}
  \caption{A $\tau$-skeleton and the threshold components exposed after lowering the threshold to $\tau'<\tau$. Bold red vertices are frontier vertices of $\mc S_\tau$; bold white vertices are frontier vertices of $\mc S_{\tau'}$. The orange boxes $C_1, C_2, C_3$ are the threshold components of $\tau, \tau'$.}
  \label{fig:threshold-skeleton-components}
\end{figure}

More generally, suppose that a connected subtree $\mc S$ has already been explored and that every vertex on its frontier has prediction at most $\tau$. For a lower threshold $\tau'$, we form a \emph{threshold component} from every frontier vertex $x$ with $\hat n_x\ge\tau'$, which is $T_{x, \tau'}:=T_x \cap \Scal_{\tau'}$, i.e., the vertices in $T_x$ with predictions at least $\tau'$. The threshold components associated with the thresholds $\tau, \tau'$ are the connected components of $\mc S_{\tau'}\setminus \mc S_{\tau}$. See \cref{fig:threshold-skeleton-components}.


\paragraph{Overview.} We use two properties of the skeletons. First, once we have explored the $\tau$-skeleton, we can complete the exploration of the entire tree in $O(\mu n/k+D+\tau)$ rounds (\cref{lem:multiplicative-euler-cleanup}). Indeed, we can take an Eulerian tour of the skeleton, regard the unexplored hanging subtrees as jobs attached to the tour, and partition the resulting ordered work as evenly as possible among the $k$ robots. The resulting load is bounded by the average predicted work $O(\mu n/k)$ plus the largest hanging-subtree prediction $O(\tau)$, while access to and from the assigned tour intervals contributes $O(D)$. In this subsection, we explore until $\tau = \Theta(\hat n_r/k) = O(\mu n/k)$, after which an Euler-tour partition completes the exploration in $O(\mu \cdot \OPT)$ time.

Second, the skeletons can be expanded iteratively and efficiently (\cref{cor:multiplicative-skeleton-expansion}). Suppose $\mc S_\tau$ has been explored, and we want to explore $\mc S_{\tau'}$ where $\tau' = \tau/2$. Every threshold component is rooted at a vertex $u$ whose prediction satisfies $n_u \le \hat n_u < \tau$, whereas each leaf $v$ in $\mc S_{\tau'}$ has $\hat n_v \ge \tau/2$ and hence $n_v \ge \tau/(2\mu)$. Hence, each component contains at most $2\mu$ leaves and consequently $O(\mu D)$ vertices (\cref{lem:multiplicative-layer}). Thus, greedily sending robots to perform DFS on the threshold components costs $O(n/k + \mu D)$, where $n/k$ is the average workload and $O(\mu D)$ is the maximum. There are $O(\log k)$ layers between the initial threshold $\tau = \hat n_r$ and the target scale $\tau =\widehat n_r/k$. This gives the simple competitive ratio $O(\mu\log k)$.

\medskip

The formal analysis begins with how to complete the exploration once $\mc S_\tau$ has been explored for some small $\tau$.

\begin{lemma}[Euler-tour cleanup]
\label{lem:multiplicative-euler-cleanup}
If we have explored $\mc S_\tau$, then the tree can be explored in $O(\mu n/k + D + \tau)$ rounds.
\end{lemma}

\begin{proof}
If $\mathcal F_\tau=\varnothing$, exploration is complete. Otherwise,
$\tau\ge1$. Take a closed depth-first walk of $\mc S_\tau$ traversing
each edge once in each direction, and record each vertex appearance; this is a doubled Euler tour of the tree.
Treat each traversal as a work item of weight one. For each
$u\in\mathcal F_\tau$, insert, at one occurrence of its parent $p(u)$,
an item of weight $2\hat n_u$ representing a depth-first exploration
of $T_u$ followed by a return to $p(u)$. This excursion takes
$2n_u\le2\hat n_u$ rounds.

Since the hanging subtrees are pairwise disjoint, the total weight is
\[
W:=2(|\mc S_\tau|-1)
   +2\sum_{u\in\mathcal F_\tau}\hat n_u
\le 2(\mu+1)n.
\]
Every item has weight at most $2\tau$. Partition the ordered items
greedily, closing each interval when its weight first reaches $W/k$
or the list ends. This produces at most $k$ intervals, each of
weight at most $W/k+2\tau$: each nonfinal interval weighs at least
$W/k$, and its last item causes an overshoot of at most $2\tau$.

Assign one robot to each interval. In parallel, each robot reaches
its first tour vertex in at most $2D$ rounds, follows its tour
segment, and performs the assigned excursions. Every hanging subtree
is explored, and the completion time is at most
\[
2D+W/k+2\tau=O(\mu n/k+D+\tau). \qedhere
\]
\end{proof}

The following lemma supports the gradual exploration of $\mc S_{\tau}$. An antichain is a set of vertices no two of which are ancestors of one another.

\begin{lemma}
\label{lem:multiplicative-antichain}
For every antichain $\mathcal A$ of vertices, the true subtree sizes of the vertices in $\mathcal A$ sum to at most $n$, and their predicted subtree sizes sum to at most $\mu n$. Consequently, at most $\mu n/\tau$ vertices in an antichain can have a prediction at least $\tau$.
\end{lemma}

\begin{proof}
The true subtrees rooted at the vertices of $\mathcal A$ are pairwise disjoint, so their sizes sum to at most $n$. The corresponding predicted sizes sum to at most $\mu n$ because $\hat n_u\leq\mu n_u$. Finally, $\hat n_u>\tau$ implies $n_u>\tau/\mu$, so fewer than $\mu n/\tau$ pairwise disjoint true subtrees can satisfy this condition.
\end{proof}

The next lemma captures the geometry of one threshold phase: a single threshold component is small.

\begin{lemma}[Threshold layers]
\label{lem:multiplicative-layer}
Suppose that we have explored $\Scal_\tau$ and are exploring $\Scal_{\tau/2}$. Then, for every frontier $x \in \Fcal_\tau$, $|T_{x, \tau/2}| \le 1+2\mu D$. 
\end{lemma}
\begin{proof}
Since $x$ belonged to the old frontier, $n_x\leq\hat n_x < \tau$. Every leaf $z$ of $\Scal_{\tau/2}$ was entered at the new threshold, and hence $n_z\geq\hat n_z/\mu\ge\tau/(2\mu)$. The true subtrees rooted at distinct leaves are pairwise disjoint and contained in $T_x$, so $T_{x,\tau/2}$ has fewer than $2\mu$ leaves, and hence at most $1+2\mu D$ vertices. 
\end{proof}

\begin{corollary}
\label{cor:multiplicative-baseline-layer}
Suppose that $\Scal_\tau$ has been explored and
$\tau\ge\hat n_r/k$. Then $\Scal_{\tau/2}$ can be explored and
all robots returned to the global root in $O(\mu D)$ additional rounds.
\end{corollary}

\begin{proof}
Return all robots to the root in $O(D)$ rounds. Whenever a robot
becomes free, send it to an unassigned threshold component to
explore it by depth-first search and return to the root.

By \cref{lem:multiplicative-layer}, each trip takes $O(\mu D)$
rounds. By \cref{lem:multiplicative-antichain}, $\Scal_{\tau/2}$
has $O(\mu n/\tau)=O(\mu k)$ leaves. Covering the components
by root-to-leaf paths therefore bounds their total size by
$O(\mu kD)$. There are also at most $O(\mu k)$ components,
so the total work, including travel, is $O(\mu kD)$.

Until the last component is assigned, all $k$ robots are busy.
Thus every component is assigned within $O(\mu D)$ rounds,
and all robots finish within another $O(\mu D)$ rounds.
\end{proof}

We can now state the baseline algorithm.

\begin{theorem}
    With predictions of multiplicative distortion $\mu$, there is a deterministic $O(\mu \log k)$-competitive algorithm.
\end{theorem}
\begin{proof}
    First, explore $\mc S_{\hat n_r}$ by a single-robot DFS (note that $\Scal_{\hat n_r}$ contains at most $\mu$ leaves and hence $O(\mu D)$ vertices). Let $B=\hat n_r/k$, let $J=\lceil\log_2 k\rceil$, and define $\tau_j=\max\{B,\hat n_r/2^j\}$ for $j=0,\ldots,J$. For each $j=1,\ldots,J$, explore the threshold components that extend $S_{\tau_{j-1}}$ to $S_{\tau_j}$ and list-schedule their depth-first searches. Consecutive thresholds differ by at most a factor of two and never fall below $B$, so \cref{cor:multiplicative-baseline-layer} gives $O(\mu D)$ rounds per phase.

    At the final threshold $\tau_J$, we have $\tau_j \le B$. Applying~\cref{lem:multiplicative-euler-cleanup} completes the cleanup in \[O(\mu n/k+D+B) = O(\mu n/k + D + \hat n/k) = O(\mu n/k+D)\] rounds. We have proved the baseline guarantee
\[
    O\!\left(\mu\frac{n}{k}+\mu D\log k\right)
    =O\!\left(\mu\log k\,\OPT\right). \qedhere
\]
\end{proof}

Roughly speaking, the total cost is $(\text{number of threshold levels}) \times (\text{cost per level})$, plus a cleanup that is usually negligible. In the baseline algorithm, $O(\log k)$ comes from the number of levels and $O(\mu D)$ is the cost per level. Our two refinements improve one bound each. Our first refinement retains the \(O(\log k)\) threshold scales but reduces the cost of each scale to \(O(\sqrt\mu\,\mathrm{OPT})\) by assigning larger teams only to components that certify correspondingly more work. Our second refinement keeps the elementary downward exploration but decreases the threshold by a quota-dependent factor, reducing the number of early scales to \(O(\mu\log\log k)\). The differences are summarized in \cref{tab:skeleton-framework}.

\begin{table}[ht]
    \centering
    \small
    \renewcommand{\arraystretch}{1.15}
    \setlength{\tabcolsep}{4pt}
    \begin{tabularx}{\textwidth}{
        @{}
        >{\raggedright\arraybackslash}p{0.25\textwidth}
        >{\raggedright\arraybackslash}X
        >{\raggedright\arraybackslash}X
        >{\centering\arraybackslash}p{0.22\textwidth}
        @{}
    }
        \toprule
        \textbf{Method}
        & \textbf{Threshold evolution}
        & \textbf{Exploration within a layer}
        & \textbf{Competitive ratio} \\
        \midrule

        Baseline (\cref{subsec:multiplicative-framework})
        & Halve each time
        & One robot per component
        & $O(\mu\log k)$ \\
        \addlinespace

        Adaptive teams (\cref{subsec:multiplicative-bootstrap})
        & Halve each time
        & Adaptive team sizes $1,2,4,\ldots,\Theta(\sqrt{\mu})$.
        & $O(\sqrt{\mu}\log k)$ \\
        \addlinespace

        Aggressive exploration (\cref{subsec:multiplicative-aggressive})
        & Quota-dependent large jumps
        & \textsc{PowerScatter} then
          adaptive tail.
        & $\widetilde O\!\left(
              \mu \log\log k
              +\mu^{3/2}
          \right)$ \\
        \bottomrule
    \end{tabularx}
    \caption{Three instantiations of the skeleton framework.}
    \label{tab:skeleton-framework}
\end{table}

\subsection{Bootstrapping component exploration}
\label{subsec:multiplicative-bootstrap}

The baseline algorithm explores each threshold component with a single robot. In this section, we provide an algorithm with an $O(\sqrt \mu \log k)$ competitive ratio by exploring the components more efficiently: we treat each component between consecutive skeletons as an unknown tree and employ known prediction-free algorithms to explore that tree. In this process, the predictions serve only to determine membership in a particular skeleton, not to explore the components between skeletons. We use the following algorithm as a black box and note that the result could be improved via the more recent $O(\log^2 k)$ guarantee in \cite{DBLP:journals/corr/abs-2609-26789}.

\begin{theorem}[Potential-function exploration \cite{cosson2024potential}]
\label{thm:multiplicative-pf}
There exist a universal constant $C$ and a deterministic algorithm with the following guarantee. Given $q\geq 1$ robots at the root of an unknown tree with $m$ vertices and depth $d$, the algorithm explores every edge and returns all robots to that root within $2m/q+O(qd)$ rounds.\footnote{Technically, \cite{cosson2024potential} is established for $q \ge 2$. However, $q=1$ is trivial, since the single-robot DFS completes the exploration in $2m+d=2m/q+O(qd)$ rounds. For brevity, we use ``the algorithm in \cite{cosson2024potential}'' to refer to the potential-function algorithm when $q \ge 2$ and DFS when $q =1$.}
\end{theorem}

\paragraph{Overview.} We assume $\mu \le k^2/4$; otherwise, a one-robot DFS is $O(k)=O(\sqrt \mu)$-competitive. Let us also assume that we have a parameter $X$ satisfying $D\leq X$, $n\leq kX$, and $X=\Theta(\OPT)$; such a parameter can be obtained through a standard guess-and-double procedure. Suppose we have an algorithm with the knowledge of $X$ that is $f(\mu, k)$-competitive. Iterate with $X_{i} = 2^i$ for $i = 1, 2,\ldots$; with each $X_i$, run the algorithm for at most $\Theta(f(\mu, k)\cdot X)$ rounds. The algorithm will complete the exploration with the first large enough guess $X_i = \Theta(\OPT)$ and the competitive ratio remains $\Theta(f(\mu, k))$. 

Our algorithm has two steps. First, we iteratively explore the skeleton until the threshold reaches $\tau^*=10\mu X$, using adaptive teams to process each threshold layer in $O(\sqrt\mu X)$ rounds. Second, we run capped one-robot probes on all remaining hanging subtrees, thereby clearing all small subtrees within $O(X \log k)$ rounds in total. Each unfinished probe certifies a subtree of true size $\Omega(X)$. Consequently, at most $k$ hanging subtrees survive, and they can again be completed using adaptive teams.

\medskip

The algorithm in \cref{alg:multiplicative-adaptive-teams} formalizes this discussion and efficiently explores at most $k$ disjoint components. It uses team sizes $1,2,4,\ldots,q_L$, where $q_L$ is the first power of two at least $\sqrt\mu$, so $q_L = \Theta(\sqrt \mu)$. At each team size $q$, the algorithm partitions the robots into $\lfloor k/q\rfloor$ teams and greedily schedules the components that survived the preceding team size, allowing each team to run a fresh copy of the algorithm from \cref{thm:multiplicative-pf} for at most $\Theta(qX)$ rounds. The following lemma establishes the correctness and the number of rounds needed for the algorithm. In the lemma, $\mc C$ can be viewed as the threshold components between $2\tau$ and $\tau$ for some $\tau \ge \Theta(\mu X)$.

\begin{algorithm}[t]
\caption{\textsc{AdaptiveTeams}$(\mathcal C,X,\mu)$}
\label{alg:multiplicative-adaptive-teams}
\begin{algorithmic}[1]
\Require The algorithm \textsc{PFExplore} in \cite{cosson2024potential}. 
\State $A \gets $ some sufficiently large constant, $L\gets\lceil\log_2\sqrt{\mu}\rceil$.
\State $\mathcal C_{\mathrm{cur}}\gets\mathcal C$.
\For{$\ell=0,1,\ldots,L$}
    \State $q\gets2^\ell$, $b\gets\lfloor k/q\rfloor$, $\mathcal C_{\mathrm{next}}\gets\varnothing$.
    \While{$\mathcal C_{\mathrm{cur}}\neq\varnothing$}
        \State Let $\mathcal B$ be any collection of at most $b$ components from $\mathcal C_{\mathrm{cur}}$.
        \State $\mathcal C_{\mathrm{cur}}\gets\mathcal C_{\mathrm{cur}}\setminus\mathcal B$.
        \ForAll{$H\in\mathcal B$ in parallel}
            \State Send $q$ robots from the global root to the root of $H$.
            \State Run \textsc{PFExplore} on $H$ with $q$ robots for at most $AqX$ rounds. 
            \If{$H$ is not completely explored}
                \State Add $H$ to $\mathcal C_{\mathrm{next}}$.
            \EndIf
            \State Return the team to the global root.
        \EndFor
    \EndWhile
    \State $\mathcal C_{\mathrm{cur}}\gets\mathcal C_{\mathrm{next}}$.
\EndFor
\State \Return \Comment{Every component has been explored.}
\end{algorithmic}
\end{algorithm}

\begin{lemma}[Adaptive teams]
\label{lem:multiplicative-adaptive-teams}
Let $\mathcal C$ be a collection of at most $k$ pairwise-disjoint rooted components, each of size at most $10 \mu X$. Then \cref{alg:multiplicative-adaptive-teams} explores every component in $\mathcal C$ in $O(\sqrt\mu\,X)$ rounds.
\end{lemma}

\begin{proof}
We first establish correctness: at the largest team size, $q_L$, the exploration always succeeds. Since each component $C$ satisfies $|C| \le 10\mu X$, running the algorithm from \cite{cosson2024potential} with $\Theta(\sqrt \mu)$ robots takes at most 
\[
\frac{20 \mu X}{\Theta(\sqrt \mu)} + \Theta(\sqrt{\mu})D = \Theta(\sqrt{ \mu} X)
\]
rounds. Taking $A$ in line 1 sufficiently large therefore ensures completion of the exploration.

It remains to bound the number of components reaching each team size. If a component reaches a team size $q\geq2$, it remained unfinished after a run with $q/2$ robots for $AqX/2$ rounds. For $q=2$, the preceding one-robot DFS directly implies that the component has size $\Omega(X)=\Omega(q^2X)$. For $q\geq4$, \cref{thm:multiplicative-pf} implies that the component has size $\Omega(q^2X)$; otherwise, the run with $q/2$ robots would have completed within its budget. Because the components are pairwise disjoint and $n\leq kX$, only $O(k/q^2)$ components reach a team size $q$. There are $\lfloor k/q\rfloor=\Omega(k/q)$ teams, so all surviving components at this team size are handled in $O(1)$ waves. Consequently, team size $q$ requires $O(qX)$ rounds. Summing $O(qX)$ over the geometrically increasing team sizes proves the lemma.
\end{proof}

\begin{corollary}[Skeleton expansion]
\label{cor:multiplicative-skeleton-expansion}
Let $\tau^*=10\mu X$. The skeleton $S_{\tau^*}$ can be explored in $O(\sqrt\mu\,X\log k)$ rounds.
\end{corollary}

\begin{proof}
Let $\tau_0=\max\{\hat n_r,\tau^*\}$. We first explore $S_{\tau_0}$ directly from the root. Every initial threshold component is rooted at a child of the root with a prediction greater than $\tau_0$. Since $n\leq \hat n_r\leq\tau_0$, each such component has at most $\mu$ leaves and at most $1+\mu D\leq2\mu X$ vertices. Moreover, the total number of leaves is less than $\mu n/\tau_0\leq k/10$, because $\tau_0\geq\tau^*=10\mu X$ and $n\leq kX$. Thus, \cref{lem:multiplicative-adaptive-teams} explores $S_{\tau_0}$ in $O(\sqrt\mu X)$ rounds.

We then use the thresholds $\tau_j=\max\{\tau^*,\tau_0/2^j\}$ until reaching $\tau^*$. Suppose $\mc S_{\tau_j}$ has been explored. We next explore $\mc S_{\tau_{j+1}}$. By \cref{lem:multiplicative-layer}, the total number of leaves $|\Lcal_{\tau_{j+1}}|$ is less than $\mu n/\tau_{j+1}\leq k/10$. Hence, there are fewer than $k$ threshold components. Each component has at most $1+2\mu D\leq3\mu X$ vertices, and the components are pairwise disjoint with total size at most $n$. Thus, we can invoke \cref{lem:multiplicative-adaptive-teams} and explore $\mc S_{\tau_{j+1}}$ in $O(\sqrt \mu X)$ rounds.
Since $\hat n_r\leq\mu n\leq\mu kX$, the number of phases is $O(1+\log(\tau_0/\tau^*))=O(\log k)$. The claimed bound follows.
\end{proof}

We next describe the cleanup. A naive Eulerian cleanup of $\mc S_{\tau^*}$ might still cost $\Theta(\mu X)$. Thus, we perform the cleanup more carefully, as described in \cref{alg:multiplicative-capped-cleanup}. 

The main goal is to clear all hanging trees with true size $n_u = O(X)$. To do so, we split the skeleton $\mc S_{\tau^*}$ into $k$ approximately equal-length intervals. The exploration proceeds in $O(\log k)$ waves. In each wave, we distribute the robots evenly among all intervals that still have hanging subtrees. The robots perform DFS on every hanging subtree in the interval, but abort after performing $\Theta(X)$ total work. Thus, either half of the robots complete $\Theta(X)$ of true work, or the number of active intervals halves. This procedure terminates after $\Theta(\log k)$ waves and takes $\Theta(\log k \cdot X)$ rounds. 

Afterward, all hanging subtrees have size $\Omega(X)$, so there are at most $O(n/X)=O(k)$ such trees, each of true size at most $O(\mu X)$. We complete the final cleanup using \cref{alg:multiplicative-adaptive-teams} in $O(\sqrt \mu X)$ rounds. 

\begin{algorithm}[ht]
\caption{\textsc{CappedProbeCleanup}$(S_{\tau^*},X,\mu)$}
\label{alg:multiplicative-capped-cleanup}
\begin{algorithmic}[1]
\State $S\gets S_{\tau^*}$ and $\mathcal F\gets\operatorname{frontier}(S)$.
\State $\mathcal H_{\mathrm{surv}}\gets\varnothing$.
\State Compute a doubled Euler tour $\Gamma$ of $S$.
\State Partition $\Gamma$ into consecutive intervals $\mc I$ of tour length at most $2X$. \Comment{$|\mc I| \le 2|S|/X\le k$.}
\ForAll{$u\in\mathcal F$}
    \State Let $p(u)$ be the parent of $u$ in $S$ and let $H_u=T_u$.
    \State Define a probe $J_u$ attached to $p(u)$ as follows:
    \Statex \qquad cross $(p(u),u)$, run DFS in $H_u$ for at most $2X$ rounds, and then return to $p(u)$.
    \State Assign $J_u$ to an interval containing one occurrence of $p(u)$ in $\Gamma$.
\EndFor
\State Let $\mathcal I_{\mathrm{act}}\subset \mc I$ be the intervals containing at least one unexecuted probe.
\While{$\mathcal I_{\mathrm{act}}\neq\varnothing$}
    \ForAll{$I\in\mathcal I_{\mathrm{act}}$ in parallel}
        \State Assign $a_I:= \lfloor k/|\mathcal I_{\mathrm{act}}| \rfloor$ robots to $I$.
        \ForAll{robots assigned to $I$ in parallel}
            \State Travel from the global root to the first tour vertex of $I$.
            \State $z\gets0$. \Comment{Finished probe work.}
            \For{each attachment encountered while scanning $I$}
                \If{it carries an unclaimed probe $J_u$ and $z<2X$}
                    \State Atomically claim $J_u$, execute $J_u$, and let $\delta_u$ be its actual duration.
                    \State \Comment{If $J_u$ is claimed, it remains claimed in future iterations.}
                    \If{$J_u$ times out}
                        \State Add $H_u$ to $\mathcal H_{\mathrm{surv}}$.
                    \EndIf
                    \State $z\gets z+\delta_u$.
                \EndIf
            \EndFor
            \State Return from the final tour vertex of $I$ to the global root.
        \EndFor
    \EndFor
    \State Remove from $\mathcal I_{\mathrm{act}}$ every interval with no unexecuted probe.
\EndWhile
\If{$\mathcal H_{\mathrm{surv}}\neq\varnothing$} 
    \State Run \textsc{AdaptiveTeams}$(\mathcal H_{\mathrm{surv}},X,\mu)$.
\EndIf
\State \Return 
\end{algorithmic}
\end{algorithm}

\begin{lemma}[Capped-probe cleanup]
\label{lem:multiplicative-capped-cleanup}
Let $\tau^*=10\mu X$. Suppose that $S_{\tau^*}$ has been explored. Then \cref{alg:multiplicative-capped-cleanup} explores the remaining tree in $O(X\log k+\sqrt\mu\,X)$ rounds.
\end{lemma}

\begin{proof}
Let us bound the time that clears all subtrees with true size at most $X$ (until line 23). We call each iteration in the while loop at line 11 an \emph{epoch}, in which we distribute the robots approximately evenly among the active intervals and allow each robot to run for $\Theta(X)$ rounds. Clearly, each epoch takes $O(X)$ rounds, and it remains to establish the number of epochs needed to complete the probing.

Let $\ell_u$ be the actual duration of the probe for $H_u$. It is easy to see that $\sum \ell_u \le \sum n_u \le n$. Since $|\mc I_{\mathrm{act}}| \le k$, we have $a_I = \lfloor k/|\mathcal I_{\mathrm{act}}| \rfloor \ge k/(2|\mc I_{\mathrm{act}}|)$. Fix an epoch. Let $\mc I_0, \mc I_1$ denote the values of the active interval set $\mc I_{\mathrm{act}}$ before and after the epoch, respectively. Two cases can occur:
\begin{enumerate}[label=(\roman*)]
    \item $|\mc I_1| \le |\mc I_0|/2$. 
    \item Otherwise, $|\mc I_1| \ge |\mc I_0|/2$. Note that every robot assigned to $I \in \mc I_1$ must complete $2X$ of probing work, since otherwise it will find an unclaimed job and execute it. Thus, the total probing work completed is at least 
    \[
    \sum_{I \in \mc I_1} a_I \cdot 2X \ge \sum_{I \in \mc I_1} \frac{k}{2|\mc I_0|} \cdot 2X = kX \cdot \frac{|\mc I_1|}{|\mc I_0|} \ge \frac{kX}{2},
    \]
\end{enumerate}
Case (i) occurs in $O(\log k)$ epochs, and case (ii) occurs in at most $\sum \ell_u /(kX/2) = O(1)$ epochs. Thus, the total number of epochs is $O(\log k)$, and the probing phase completes in $O(X \log k)$ rounds.

After the probing phase, every surviving hanging subtree has true size greater than $X$ (since otherwise a probe would have completed it). Since the survivors are pairwise disjoint and their total size is at most $n\leq kX$, at most $k$ of them survive. In addition, every survivor satisfies $n_u\leq\hat n_u\leq\tau^*=10\mu X$. We can therefore apply \cref{lem:multiplicative-adaptive-teams} to complete all survivors in $O(\sqrt\mu\,X)$ rounds. The total cleanup time is $O(X\log k+\sqrt\mu\,X)$.
\end{proof}

Thus, \cref{cor:multiplicative-skeleton-expansion} explores $S_{10\mu X}$ in $O(\sqrt\mu\,X\log k)$ rounds, and \cref{lem:multiplicative-capped-cleanup} completes the remaining exploration in $O(X\log k+\sqrt\mu\,X)$ rounds. Together, they give the following accelerated algorithm.

\begin{theorem}
\label{thm:multiplicative-bootstrap}
With predictions of multiplicative distortion $\mu$, there is a deterministic $O(\sqrt\mu\log k)$-competitive algorithm.
\end{theorem}

\subsection{Aggressive exploration for small $\mu$}
\label{subsec:multiplicative-aggressive}

Next, we introduce the second refinement, which gives an $O(\mu \log \log k + \mu^{1.5})$ competitive ratio. This refinement reduces the number of relevant scales by decreasing the threshold much more aggressively during the initial part of the exploration, particularly when $\mu$ is small. Henceforth, assume that $\mu\leq k$, since otherwise an ordinary one-robot DFS is $O(k)=O(\mu)$-competitive. 

\paragraph{Overview.} Intuitively, if $\mu$ is small, we should be able to explore a much larger portion of most early skeletons and decrease $\tau$ much more aggressively. It may be helpful to think about the case in which $\mu$ is a constant, such as $\mu=2$. 

Let $U = \mu\widehat n_r/k$ be an estimated quota unit, i.e., roughly one robot should be assigned per $U$ vertices. We assign each frontier root $u$ a quota of $k_u=\lfloor\widehat s(u)/U\rfloor$ robots. Rather than lowering every threshold by the fixed factor $2$, we choose a task-dependent reduction factor determined by its quota. In each frontier subtree, we simply distribute the robots among eligible children proportionally to their predicted subtree sizes. We show that a frontier root with quota $q$ generates frontier vertices whose quotas are at most $O(\mu q^{1-1/\mu})$. Consequently, along every chain of successively generated tasks, the quota falls below $(64\mu)^\mu$ after $O(\mu\log\log k)$ global levels. Each global level consists of downward exploration followed by a return to the global root and therefore takes $O(D)$ rounds. Hence, we reveal the skeleton with \[\tau_0 = (64\mu)^\mu \cdot U =(64\mu)^\mu\cdot \mu \hat n_r/k \le (64\mu)^\mu \mu^2 X.\] Afterward, we reuse the bootstrapped algorithm in the previous section to explore until $\tau=\mu X$ in $O(\mu \log \mu)$ phases, each costing $O(\sqrt \mu X)$, and eventually perform the standard Euler-tour cleanup in $O(\mu X)$ rounds. 

\medskip

To formalize this sketch, consider how to explore a single frontier task. Suppose that a task is rooted at $u$ and receives $m$ robots. For a reduction factor $R$ to be chosen later, we explore the local skeleton down to threshold $\tau=\hat n_u/R$ using the procedure \textsc{PowerScatter} in \cref{alg:power-scatter}. The procedure distributes robots among eligible children proportionally to their predicted subtree sizes, rounding down the allocation to every child except the child with the maximum predicted share, which receives all remaining robots.

\begin{algorithm}[ht]
\caption{\textsc{PowerScatter}\((u,k_u,\tau)\)}
\label{alg:power-scatter}
\begin{algorithmic}[1]
\State Reveal the children of $u$
\State $H(u)\gets\{v\in\mathrm{ch}(u):\hat n_v>\tau\}$
\State Declare every child in $\mathrm{ch}(u)\setminus H(u)$ to be a boundary task
\If{$H(u)=\varnothing$}
    \State Stop all $k_u$ robots at $u$ and \Return
\EndIf
\State $\hat \alpha_v\gets\hat n_v/\sum_{v\in H(u)}\hat n_v$ for every $v\in H(u)$
\State Choose $h\in\arg\max_{v\in H(u)}\alpha_v$.
\ForAll{$v\in H(u)\setminus\{h\}$}
    \State $k_v\gets\lfloor\alpha_v k_u\rfloor$.
\EndFor
\State $k_h\gets k_u-\sum_{v\in H(u)\setminus\{h\}}k_v$.
\ForAll{$v\in H(u)$ in parallel}
    \State Send the $k_v$ robots to $v$.
    \State Run \textsc{PowerScatter}\((v,k_v,\tau)\).
\EndFor
\end{algorithmic}
\end{algorithm}

The following lemma gives a sufficient condition for \textsc{PowerScatter} to reach the entire local skeleton and determines the choice of $R$.

\begin{lemma}[Power scattering]
\label{lem:multiplicative-power-scatter}
If $k_u\geq4(\mu R)^\mu$, then \textnormal{\textsc{PowerScatter}}\((u,k_u,\hat n_u/R)\) reaches every vertex of the local $\hat n_u/R$-skeleton rooted at $u$.
\end{lemma}

\begin{proof}
Fix a path $u=x_0,x_1,\ldots,x_t=z$ in the local skeleton, and let $\hat\alpha_i$ be the fraction of robots assigned at $x_i$ to $x_{i+1}$. Let $A_i=\sum_{v\in H(x_i)\setminus\{x_{i+1}\}}n_v$ be the total true size of the other eligible child subtrees. Then
\[
    \hat\alpha_i
    \geq \frac{s(x_{i+1})}{s(x_{i+1})+\mu A_i}
    \geq \left(\frac{s(x_{i+1})}{s(x_{i+1})+A_i}\right)^\mu
    \geq \left(\frac{s(x_{i+1})}{s(x_i)}\right)^\mu,
\]
where the second inequality uses $a/(a+\mu b)\geq(a/(a+b))^\mu$ for $a,b\geq0$ and $\mu\geq1$, and the last inequality uses $s(x_i)\geq s(x_{i+1})+A_i$.

Let $Z_0=k_u$ and $Z_i=k_u\prod_{j=0}^{i-1}\hat\alpha_j$ for $i\geq1$. By telescoping, $Z_i\geq k_u(s(x_i)/n_u)^\mu$. Let $k_{i}$ be the actual number of robots reaching $x_i$. We claim that $k_{i}\geq Z_i-2$ for every $i$. The claim is immediate for $i=0$. Suppose it holds for $i$. \begin{itemize}
    \item If $\hat\alpha_i>1/2$, then $x_{i+1}$ is the unique heaviest eligible child and receives the remainder, so \[k_{i+1}\geq\hat \alpha_i k_i\geq Z_{i+1}-2.\]
    \item If $\hat\alpha_i\leq1/2$, then either the flooring rule or the remainder rule gives \[k_{i+1}\geq\hat\alpha_i k_i-1\geq Z_{i+1}-(2\hat\alpha_i+1)\geq Z_{i+1}-2.\] 
\end{itemize} 
Thus, the rounding loss remains bounded by two rather than accumulating with the depth. Every non-root vertex $z$ in the local skeleton satisfies $\hat n_z>\hat n_u/R$, and therefore $n_z>n_u/(\mu R)$. It follows that $Z_t>m/(\mu R)^\mu\geq4$, so $r_t\geq Z_t-2>2$. In particular, at least one robot reaches $z$. Since the path was arbitrary, \textsc{PowerScatter} reaches every vertex of the local skeleton.
\end{proof}

The aggressive exploration phase proceeds as follows. Let $U=\mu \hat n_r/k$ be the quota unit, and assign every frontier task $u$ the quota $k_u=\lfloor\hat n_u/U\rfloor=\lfloor k\hat n_u/(\mu \hat n_r)\rfloor$. At every global level, the current frontier tasks form an antichain. Therefore, by \cref{lem:multiplicative-antichain}, \[\sum_u k_u\leq U^{-1}\sum_u\hat n_u\leq\mu n/(\mu \hat n_r/k)=kn/\hat n_r\leq k.\] Hence all aggressive tasks at one global level can receive their quotas simultaneously; any unused robots remain at the global root.

Set $M_\mu=\lceil(64\mu)^\mu\rceil$. A task $u$ is \emph{aggressive} if $k_u\geq M_\mu$. For such a task, set $R_u:=\lfloor k_u^{1/\mu}/(4\mu)\rfloor$. The choice of $M_\mu$ ensures that $R_u\geq1$, and the definition gives $k_u\geq4(\mu R_u)^\mu$. We send $k_u$ robots from the global root to $u$, run \textsc{PowerScatter} with local threshold $\hat n_u/R_u$, and then return all of these robots to the global root. The boundary tasks generated by the call become candidates for the next global level. Nonaggressive tasks are deferred to the final frontier. Since every robot travels only along root-to-vertex paths of length at most $D$, one global level takes $O(D)$ rounds.

\begin{algorithm}[ht]
\caption{Aggressive Exploration Phase}
\label{alg:multiplicative-exploration-aggressive}
\begin{algorithmic}[1]
\State Reveal the children of $r$.
\State $\mathcal F\gets\mathrm{ch}(r)$, $\mathcal B\gets\varnothing$, $M_\mu\gets\lceil(64\mu)^\mu\rceil$.
\While{$\mathcal F\neq\varnothing$}
    \State $\mathcal F'\gets\varnothing$.
    \ForAll{$u\in\mathcal F$ in parallel}
        \State $k_u\gets\lfloor k\hat n_u/(\mu \hat n_r)\rfloor$.
        \If{$k_u\geq M_\mu$}
            \State $R_u\gets\lfloor k_u^{1/\mu}/(4\mu)\rfloor$.
            \State Send $k_u$ robots from $r$ to $u$ and run \textsc{PowerScatter}\((u,k_u,\hat n_u/R_u)\).
            \State Add every boundary task generated by this call to $\mathcal F'$.
        \Else
            \State Add $u$ to $\mathcal B$.
        \EndIf
    \EndFor
    \State Return all robots to $r$.
    \State $\mathcal F\gets\mathcal F'$.
\EndWhile
\State \Return $\mathcal B$.
\end{algorithmic}
\end{algorithm}

\begin{lemma}
\label{lem:multiplicative-aggressive-levels}
The aggressive exploration phase has $O(1+\mu\log\log k)$ global levels and therefore completes in $O((1+\mu\log\log k)D)$ rounds.
\end{lemma}

\begin{proof}
The quota decreases rapidly along every chain of successively generated aggressive tasks. Suppose that $v$ is a boundary task generated while processing $u$. Then $\hat n_v\leq\hat n_u/R_u$, and hence
\[
    k_v
    \leq \frac{\hat n_v}{U}
    \leq \frac{\hat n_u}{UR_u}
    < \frac{k_u+1}{R_u}.
\]
Because $k_u\geq M_\mu$, we have $R_u\geq k_u^{1/\mu}/(8\mu)$ and $k_u+1\leq2k_u$. Therefore,
\[
    k_v\leq16\mu k_u^{1-1/\mu}.
\]

Along a chain of successively generated aggressive tasks, let $m_j$ be the quota at global level $j$. We have $m_0\leq k$ and $m_{j+1}\leq16\mu m_j^{1-1/\mu}$. When $\mu=1$, the next quota is at most $16<M_\mu$, so the chain ends after one level. For $\mu>1$, put $p=1-1/\mu$. Taking logarithms and iterating gives
\[
    \log m_j
    \leq p^j\log k+\mu\log(16\mu),
    \qquad
    p^j\leq e^{-j/\mu}.
\]
After $O(1+\mu\log\log k)$ levels, the first term is at most $\mu\log2$. Consequently, $\log m_j<\mu\log(64\mu)\leq\log M_\mu$, so the task is no longer aggressive. Since every global level takes $O(D)$ rounds, the claimed completion time follows.
\end{proof}

At the end of the aggressive phase, every deferred frontier task $u$ satisfies $k_u<M_\mu$ and hence $\hat n_u<M_\mu U$. Thus the explored subtree contains the skeleton at threshold $\tau_0=M_\mu U=M_\mu\mu \hat n_r/k$, although it may contain additional vertices. We now bridge the remaining gap to threshold $O(\mu X)$ using the bootstrapped layer procedure.

\begin{theorem}[Aggressive skeletons]
\label{thm:multiplicative-aggressive}
With predictions of multiplicative distortion $\mu$, there is a deterministic algorithm with competitive ratio $O(\mu(1+\log\log k)+\mu^{3/2}\log\mu)$.
\end{theorem}

\begin{proof}
If $\mu>k$, a one-robot DFS is $O(k)=O(\mu)$-competitive, so assume $\mu\leq k$. First run the aggressive exploration phase. By \cref{lem:multiplicative-aggressive-levels} and $D\leq X$, this costs $O(\mu(1+\log\log k)X)$ rounds.

Let $\tau_0=M_\mu\mu \hat n_r/k$. Since $\hat n_r\leq\mu n\leq\mu kX$, we have $\tau_0\leq M_\mu\mu^2X$. If $\tau_0>20\mu X$, repeatedly halve the threshold while the new threshold remains at least $10\mu X$; otherwise, skip this step. Each threshold phase costs $O(\sqrt\mu X)$ rounds by the same analysis as in \cref{cor:multiplicative-skeleton-expansion}. In either case, we end with a threshold of at most $20\mu X$. The number of phases is $O(1+\log(\tau_0/(\mu X)))=O(\mu\log\mu)$ because $\log M_\mu=O(\mu\log\mu)$. The total cost of this dyadic tail is therefore $O(\mu^{3/2}\log\mu\,X)$.

We are left with an explored subtree whose frontier predictions are $O(\mu X)$. Applying the Euler-tour cleanup from \cref{lem:multiplicative-euler-cleanup} costs $O(\mu X)$ rounds. Since $X=\Theta(\OPT)$, the total competitive ratio is $O(\mu(1+\log\log k)+\mu^{3/2}\log\mu)$.
\end{proof}

\subsection{Unknown prediction quality}
\label{subsec:multiplicative-assumption}
Finally, we remove the assumption that the prediction quality is known by reducing the unknown-quality setting to the known-quality setting. We guess the expansion and contraction simultaneously and combine all guesses using a countable weighted version of the min-operator.

\begin{lemma}[Weighted min-operator]
\label{lem:weighted-min-operator}
Let $\mathcal I$ be a countable index set, and let
$\{w_a\}_{a\in\mathcal I}$ be positive weights satisfying
$\sum_{a\in\mathcal I}w_a\leq 1$. For every $a\in\mathcal I$, let
$\mathcal A_a$ be a deterministic exploration procedure that starts with all
robots at the root. There is a deterministic master algorithm with the
following property. If, on the given instance, some procedure
$\mathcal A_{a^\star}$ completes the exploration and returns all robots to the
root within $T^\star$ rounds, then the master algorithm completes within
$O(T^\star/w_{a^\star})$ rounds. The master algorithm need not know either
$a^\star$ or $T^\star$.
\end{lemma}

\begin{proof}
For $h=0,1,2,\ldots$, let the nominal length of epoch $h$ be $H_h=2^h$.
During epoch $h$, consider, in an arbitrary fixed order, every index $a$
satisfying $w_aH_h\geq 1$. For each such $a$, start a fresh virtual execution of
$\mathcal A_a$ and simulate it for
$b_{h,a}=\lfloor w_aH_h\rfloor$ rounds, or until it completes. If the
simulation does not complete, return all robots to the root before starting
the next simulation.

There are only finitely many indices satisfying $w_aH_h\geq 1$, since
otherwise their weights would sum to more than one. Moreover, the total
simulation time in epoch $h$ is at most
$\sum_{a\in\mathcal I}b_{h,a}\leq H_h\sum_{a\in\mathcal I}w_a\leq H_h$.
Returning the robots to the root after each simulation costs at most the
length of that simulation. Thus, epoch $h$ costs at most $2H_h$ rounds.

Let $h^\star$ be the first epoch satisfying
$w_{a^\star}H_{h^\star}\geq 2(T^\star+1)$. Then
$b_{h^\star,a^\star}\geq T^\star$, so the fresh execution of
$\mathcal A_{a^\star}$ completes during epoch $h^\star$. Since the epoch
lengths grow geometrically, the total cost through epoch $h^\star$ is
$O(H_{h^\star})=O((T^\star+1)/w_{a^\star})$. On every nontrivial instance,
$T^\star\geq 1$, proving the claimed bound.
\end{proof}

The reduction is proved next. 

\begin{lemma}
\label{lem:multiplicative-reduction}
    Suppose that there is an algorithm that, given normalized predictions, has running time $f(\mu, k) \cdot \OPT$, where $f(\cdot,k)$ is nondecreasing. Then there is an algorithm that does not know $\mu$ and has running time $f(4\mu, k)\cdot O(\log^3(2+\mu)) \cdot \OPT$.
\end{lemma}

\begin{proof}
Write $\mu_e = \max_{u} \hat n_u / n_u$ and $\mu_c = \max_u n_u / \hat n_u$ for the contraction and expansion factors, respectively, so $\mu = \mu_e \mu _c$. For every pair of integers $i, j \ge 0$, define a candidate $(i, j)$ that posits an expansion $\mu_e \le 2^i$ and a contraction $\mu_c \le 2^j$. The candidate $(i, j)$ forms the predictions $p_{i, j}(u) = 2^{j} \cdot \hat n_u$ and runs the one-sided algorithm on these predictions.

Assign the candidate $(i,j)$ the weight $w_{i,j}=1/(2(i+j+1)^3)$. These weights are summable, since
\[
    \sum_{i=0}^{\infty}\sum_{j=0}^{\infty}w_{i,j}
    =\frac{1}{2}\sum_{i+j=0}^{\infty}\frac{i+j+1}{(i+j+1)^3}
    =\frac{1}{2}\sum_{i+j=0}^{\infty}\frac{1}{(i+j+1)^2}
    <1.
\]
We combine all candidates using Lemma~\ref{lem:weighted-min-operator}.

It remains to identify a valid candidate. Let $i^\star=\lceil\log_2\mu_e\rceil$ and $j^\star=\lceil\log_2\mu_c\rceil$. Both $i^\star$ and $j^\star$ are nonnegative. Consider the candidate $(i^\star,j^\star)$. For every vertex $u$, we have $p_{i^\star,j^\star}(u)=2^{j^\star}\hat n_u\geq\mu_c\hat n_u\geq n_u$. Moreover, $p_{i^\star,j^\star}(u)=2^{j^\star}\hat n_u\leq2^{j^\star}\mu_e n_u\leq2^{i^\star+j^\star}n_u$. Thus, the candidate $(i^\star,j^\star)$ satisfies the one-sided promise with distortion parameter $2^{i^\star+j^\star}$. This guessed distortion is within a constant factor of the true distortion, since $2^{i^\star+j^\star}\leq4\mu_e\mu_c=4\mu$. Furthermore, $i^\star+j^\star\leq\log_2\mu_e+\log_2\mu_c+2=\log_2\mu+2$.

The standalone completion time of this valid candidate is at most \[T^\star\leq C f(2^{i^\star+j^\star},k)\OPT\leq C f(4\mu,k)\OPT\] for some universal constant $C$. Its inverse weight satisfies $1/w_{i^\star,j^\star}=2(i^\star+j^\star+1)^3=O(\log^3(2+\mu))$. Applying Lemma~\ref{lem:weighted-min-operator}, the total simulation time is therefore $O(f(4\mu,k)\log^3(2+\mu)\OPT)$.
\end{proof}

Applying \cref{lem:multiplicative-reduction} to \cref{thm:multiplicative-bootstrap} and \ref{thm:multiplicative-aggressive} gives \cref{thm:multiplicative}.

\section{Conclusion and Discussion} 
We initiated the study of collective tree exploration with predicted subtree sizes. This form of advice is deliberately weaker than a predicted map: it does not reveal the topology below a frontier, but only estimates the amount of work hidden there, and the available information depends on the exploration completed so far. Nevertheless, exact subtree sizes already suffice to recover the offline benchmark $O(n/k+D)$. This shows that, for the Euler-tour benchmark, accurate workload information can substitute for advance knowledge of the entire tree.

Under the maximum additive error model, our algorithm completes the exploration in $O(\OPT+E)$ rounds, and our lower bound rules out any uniform $O(\OPT)+o(E)$ guarantee for deterministic algorithms. Under the multiplicative distortion model, we obtain an $\widetilde O(\min\{\mu\log\log k+\mu^{3/2},\sqrt{\mu}\log k\})$-competitive algorithm. These results leave several natural questions open.

\begin{enumerate}
    \item The dependence on the multiplicative distortion is unlikely to be final. In particular, when $\mu=1$, the predictions are exact up to a common rescaling, and an $O(1)$ competitive ratio is achievable. It would be particularly interesting to obtain an $f(\mu)$-competitive algorithm whose ratio depends only on $\mu$.

    \item Our additive guarantee hides a constant factor multiplying $\OPT$. It would be interesting to determine whether one can obtain the finer-grained bound $2n/k+O(D+E)$, paying only an additive penalty for the depth and the prediction error.

    \item One can also consider alternative error measures. For example, let the \emph{total additive error} be $E_1=\sum_u|\hat n_u-n_u|$. Can errors distributed across different subtrees be amortized to obtain a completion time of the form $O(\OPT+f(n, k) \cdot E_1)$? Note that obtaining such a result for $f(n, k) = 1/n$ would simultaneously yield an $O(\mu)$-competitive algorithm for the multiplicative distortion model and an $O(\OPT+E)$ guarantee for the maximum additive error model.

    \item Stochastic models for the predictions are also interesting. For example, the predictions may be unbiased, independent across suitable regions, or satisfy concentration guarantees rather than deterministic worst-case error bounds. Such assumptions may permit guarantees that depend on the variance or tail behavior of the prediction errors rather than their worst-case magnitude.

    \item Finally, it would be interesting to study collective tree exploration with predictions in the \emph{asynchronous} model (see \cite{cosson2024potential}). In this setting, predictions may need to compensate not only for uncertainty about the remaining work, but also for robot failures. 
\end{enumerate}

More broadly, collective tree exploration highlights a feature that is largely absent from single-agent learning-augmented problems: inaccurate predictions affect not only where the algorithm should act, but also how a shared pool of robots should be allocated among many competing tasks. We hope that the framework developed here opens further research directions for prediction-augmented parallel and distributed algorithms, in which the effects of prediction error on both allocation and coordination are accounted for explicitly.

\paragraph{AI Disclosure} ChatGPT 5.6-Sol Pro was used for brainstorming proof strategies, verifying proposed algorithms, and proofreading and polishing the manuscript. The algorithm ideas were proposed by the authors, and the AI assisted with developing some details and calculating exact bounds. The authors verified the correctness and originality of all content.

\printbibliography

@article{fraigniaud2006collective,
  author  = {Fraigniaud, Pierre and G{\k{a}}sieniec, Leszek and Kowalski, Dariusz R. and Pelc, Andrzej},
  title   = {Collective Tree Exploration},
  journal = {Networks},
  volume  = {48},
  number  = {3},
  pages   = {166--177},
  year    = {2006},
  doi     = {10.1002/net.20127},
  url     = {https://doi.org/10.1002/net.20127}
}

@inproceedings{dynia2007why,
  author    = {Dynia, Miroslaw and {\L}opusza{\'n}ski, Jakub and Schindelhauer, Christian},
  title     = {Why Robots Need Maps},
  booktitle = {Structural Information and Communication Complexity},
  series    = {Lecture Notes in Computer Science},
  volume    = {4474},
  pages     = {41--50},
  publisher = {Springer},
  year      = {2007},
  doi       = {10.1007/978-3-540-72951-8_5},
  url       = {https://doi.org/10.1007/978-3-540-72951-8_5}
}

@inproceedings{SaroyaEtAl2020TunnelPredictions,
  author    = {Saroya, Manish and Best, Graeme and Hollinger, Geoffrey A.},
  title     = {Online Exploration of Tunnel Networks Leveraging
               Topological {CNN}-Based World Predictions},
  booktitle = {2020 IEEE/RSJ International Conference on
               Intelligent Robots and Systems (IROS)},
  pages     = {6038--6045},
  year      = {2020},
  doi       = {10.1109/IROS45743.2020.9341170},
  url       = {https://doi.org/10.1109/IROS45743.2020.9341170}
}

@inproceedings{azar2021flow,
  title={Flow time scheduling with uncertain processing time},
  author={Azar, Yossi and Leonardi, Stefano and Touitou, Noam},
  booktitle={proceedings of the 53rd Annual ACM SIGACT Symposium on Theory of Computing},
  pages={1070--1080},
  year={2021}
}

@article{antoniadis2023online,
  title={Online metric algorithms with untrusted predictions},
  author={Antoniadis, Antonios and Coester, Christian and Eli{\'a}{\v{s}}, Marek and Polak, Adam and Simon, Bertrand},
  journal={ACM transactions on algorithms},
  volume={19},
  number={2},
  pages={1--34},
  year={2023},
  publisher={ACM, New York, NY}
}

@article{blumofe1999scheduling,
  title={Scheduling multithreaded computations by work stealing},
  author={Blumofe, Robert D and Leiserson, Charles E},
  journal={Journal of the ACM (JACM)},
  volume={46},
  number={5},
  pages={720--748},
  year={1999},
  publisher={ACM, New York, NY, USA}
}

@inproceedings{gupta2026optimal,
  title={An Optimal Online Algorithm for Robust Flow Time Scheduling},
  author={Gupta, Anupam and Kumar, Amit and Panigrahi, Debmalya and Wang, Zhaozi},
  booktitle={Proceedings of the 2026 Annual ACM-SIAM Symposium on Discrete Algorithms (SODA)},
  pages={1214--1238},
  year={2026},
  organization={SIAM}
}

@article{TanEtAl2026FourCNet,
  author  = {Tan, Aaron Hao and Narasimhan, Siddarth and Nejat, Goldie},
  title   = {{4CNet}: A Diffusion Approach to Map Prediction for
             Decentralized Multirobot Exploration},
  journal = {IEEE Transactions on Robotics},
  volume  = {42},
  pages   = {1371--1388},
  year    = {2026},
  doi     = {10.1109/TRO.2026.3666133},
  url     = {https://doi.org/10.1109/TRO.2026.3666133}
}

@article{LupertoEtAl2026MapCompleteness,
  author  = {Luperto, Matteo and Ferrara, Marco Maria and
             Princisgh, Matteo and Boracchi, Giacomo and
             Amigoni, Francesco},
  title   = {Estimating Map Completeness in Robot Exploration},
  journal = {Autonomous Robots},
  volume  = {50},
  year    = {2026},
  note    = {Article 6},
  doi     = {10.1007/s10514-025-10221-8},
  url     = {https://doi.org/10.1007/s10514-025-10221-8}
}

@inproceedings{dynia2006smart,
  author    = {Dynia, Miroslaw and Kuty{\l}owski, Jaroslaw and Meyer auf der Heide, Friedhelm and Schindelhauer, Christian},
  title     = {Smart Robot Teams Exploring Sparse Trees},
  booktitle = {Mathematical Foundations of Computer Science 2006},
  series    = {Lecture Notes in Computer Science},
  volume    = {4162},
  pages     = {327--338},
  publisher = {Springer},
  year      = {2006},
  doi       = {10.1007/11821069_29},
  url       = {https://doi.org/10.1007/11821069_29}
}

@inproceedings{dynia2006power,
  author    = {Dynia, Miroslaw and Korzeniowski, Miroslaw and Schindelhauer, Christian},
  title     = {Power-Aware Collective Tree Exploration},
  booktitle = {Architecture of Computing Systems -- {ARCS} 2006},
  series    = {Lecture Notes in Computer Science},
  volume    = {3894},
  pages     = {341--351},
  publisher = {Springer},
  year      = {2006},
  doi       = {10.1007/11682127_24},
  url       = {https://doi.org/10.1007/11682127_24}
}

@article{dereniowski2015fast,
  author  = {Dereniowski, Dariusz and Disser, Yann and Kosowski, Adrian and Paj{\k{a}}k, Dominik and Uzna{\'n}ski, Przemys{\l}aw},
  title   = {Fast Collaborative Graph Exploration},
  journal = {Information and Computation},
  volume  = {243},
  pages   = {37--49},
  year    = {2015},
  doi     = {10.1016/j.ic.2014.12.005},
  url     = {https://doi.org/10.1016/j.ic.2014.12.005}
}

@article{disser2020general,
  author  = {Disser, Yann and Mousset, Frank and Noever, Andreas and {\v{S}}kori{\'c}, Nemanja and Steger, Angelika},
  title   = {A General Lower Bound for Collaborative Tree Exploration},
  journal = {Theoretical Computer Science},
  volume  = {811},
  pages   = {70--78},
  year    = {2020},
  doi     = {10.1016/j.tcs.2018.03.006},
  url     = {https://doi.org/10.1016/j.tcs.2018.03.006}
}

@article{brass2011multirobot,
  author  = {Brass, Peter and Cabrera-Mora, Flavio and Gasparri, Andrea and Xiao, Jizhong},
  title   = {Multirobot Tree and Graph Exploration},
  journal = {IEEE Transactions on Robotics},
  volume  = {27},
  number  = {4},
  pages   = {707--717},
  year    = {2011},
  doi     = {10.1109/TRO.2011.2121170},
  url     = {https://doi.org/10.1109/TRO.2011.2121170}
}

@inproceedings{ortolf2014recursive,
  author    = {Ortolf, Christian and Schindelhauer, Christian},
  title     = {A Recursive Approach to Multi-Robot Exploration of Trees},
  booktitle = {Structural Information and Communication Complexity},
  series    = {Lecture Notes in Computer Science},
  volume    = {8576},
  pages     = {343--354},
  publisher = {Springer},
  year      = {2014},
  doi       = {10.1007/978-3-319-09620-9_26},
  url       = {https://doi.org/10.1007/978-3-319-09620-9_26}
}

@inproceedings{cosson2024potential,
  author    = {Cosson, Romain and Massouli{\'e}, Laurent},
  title     = {Collective Tree Exploration via Potential Function Method},
  booktitle = {15th Innovations in Theoretical Computer Science Conference ({ITCS} 2024)},
  series    = {Leibniz International Proceedings in Informatics (LIPIcs)},
  volume    = {287},
  pages     = {35:1--35:22},
  publisher = {Schloss Dagstuhl--Leibniz-Zentrum f{\"u}r Informatik},
  year      = {2024},
  doi       = {10.4230/LIPIcs.ITCS.2024.35},
  url       = {https://doi.org/10.4230/LIPIcs.ITCS.2024.35}
}

@inproceedings{cosson2024treemining,
  author    = {Cosson, Romain},
  title     = {Breaking the $k/\log k$ Barrier in Collective Tree Exploration via Tree-Mining},
  booktitle = {Proceedings of the 2024 Annual ACM-SIAM Symposium on Discrete Algorithms ({SODA})},
  pages     = {4264--4282},
  publisher = {SIAM},
  year      = {2024},
  doi       = {10.1137/1.9781611977912.148},
  url       = {https://doi.org/10.1137/1.9781611977912.148}
}

@inproceedings{lykouris2018competitive,
  author    = {Lykouris, Thodoris and Vassilvitskii, Sergei},
  title     = {Competitive Caching with Machine Learned Advice},
  booktitle = {Proceedings of the 35th International Conference on Machine Learning},
  series    = {Proceedings of Machine Learning Research},
  volume    = {80},
  pages     = {3296--3305},
  publisher = {PMLR},
  year      = {2018},
  url       = {https://proceedings.mlr.press/v80/lykouris18a.html}
}

@inproceedings{purohit2018improving,
  author    = {Purohit, Manish and Svitkina, Zoya and Kumar, Ravi},
  title     = {Improving Online Algorithms via {ML} Predictions},
  booktitle = {Advances in Neural Information Processing Systems 31},
  pages     = {9684--9693},
  publisher = {Curran Associates, Inc.},
  year      = {2018},
  url       = {https://proceedings.neurips.cc/paper/2018/hash/73a427badebe0e32caa2e1fc7530b7f3-Abstract.html}
}

@incollection{mitzenmacher2021algorithms,
  author    = {Mitzenmacher, Michael and Vassilvitskii, Sergei},
  title     = {Algorithms with Predictions},
  booktitle = {Beyond the Worst-Case Analysis of Algorithms},
  editor    = {Roughgarden, Tim},
  pages     = {646--662},
  publisher = {Cambridge University Press},
  year      = {2021},
  doi       = {10.1017/9781108637435.037},
  url       = {https://doi.org/10.1017/9781108637435.037}
}

@inproceedings{azar2022distortion,
  author    = {Azar, Yossi and Peretz, Eldad and Touitou, Noam},
  title     = {Distortion-Oblivious Algorithms for Scheduling on Multiple Machines},
  booktitle = {33rd International Symposium on Algorithms and Computation ({ISAAC} 2022)},
  series    = {Leibniz International Proceedings in Informatics (LIPIcs)},
  volume    = {248},
  pages     = {16:1--16:18},
  publisher = {Schloss Dagstuhl--Leibniz-Zentrum f{\"u}r Informatik},
  year      = {2022},
  doi       = {10.4230/LIPIcs.ISAAC.2022.16},
  url       = {https://doi.org/10.4230/LIPIcs.ISAAC.2022.16}
}

@article{eberle2022robustification,
  author    = {Eberle, Franziska and Lindermayr, Alexander and Megow, Nicole and N{\"o}lke, Lukas and Schl{\"o}ter, Jens},
  title     = {Robustification of Online Graph Exploration Methods},
  journal   = {Proceedings of the AAAI Conference on Artificial Intelligence},
  volume    = {36},
  number    = {9},
  pages     = {9732--9740},
  year      = {2022},
  doi       = {10.1609/aaai.v36i9.21208},
  url       = {https://doi.org/10.1609/aaai.v36i9.21208}
}

@inproceedings{banerjee2023searching,
  author    = {Banerjee, Siddhartha and Cohen-Addad, Vincent and Gupta, Anupam and Li, Zhouzi},
  title     = {Graph Searching with Predictions},
  booktitle = {14th Innovations in Theoretical Computer Science Conference ({ITCS} 2023)},
  series    = {Leibniz International Proceedings in Informatics (LIPIcs)},
  volume    = {251},
  pages     = {12:1--12:24},
  publisher = {Schloss Dagstuhl--Leibniz-Zentrum f{\"u}r Informatik},
  year      = {2023},
  doi       = {10.4230/LIPIcs.ITCS.2023.12},
  url       = {https://doi.org/10.4230/LIPIcs.ITCS.2023.12}
}

@inproceedings{bernardini2022universal,
  author    = {Bernardini, Giulia and Lindermayr, Alexander and Marchetti-Spaccamela, Alberto and Megow, Nicole and Stougie, Leen and Sweering, Michelle},
  title     = {A Universal Error Measure for Input Predictions Applied to Online Graph Problems},
  booktitle = {Advances in Neural Information Processing Systems 35},
  pages     = {3178--3190},
  publisher = {Curran Associates, Inc.},
  year      = {2022},
  doi       = {10.52202/068431-0230},
  url       = {https://proceedings.neurips.cc/paper_files/paper/2022/hash/15212bd2265c4a3ab0dbc1b1982c1b69-Abstract-Conference.html}
}

@inproceedings{yamauchi1997frontier,
  author    = {Yamauchi, Brian},
  title     = {A Frontier-Based Approach for Autonomous Exploration},
  booktitle = {Proceedings of the 1997 IEEE International Symposium on Computational Intelligence in Robotics and Automation},
  pages     = {146--151},
  publisher = {IEEE},
  year      = {1997},
  doi       = {10.1109/CIRA.1997.613851},
  url       = {https://doi.org/10.1109/CIRA.1997.613851}
}

@article{burgard2005coordinated,
  author  = {Burgard, Wolfram and Moors, Mark and Stachniss, Cyrill and Schneider, Frank E.},
  title   = {Coordinated Multi-Robot Exploration},
  journal = {IEEE Transactions on Robotics},
  volume  = {21},
  number  = {3},
  pages   = {376--386},
  year    = {2005},
  doi     = {10.1109/TRO.2004.839232},
  url     = {https://doi.org/10.1109/TRO.2004.839232}
}

@article{osswald2016speeding,
  author  = {O{\ss}wald, Stefan and Bennewitz, Maren and Burgard, Wolfram and Stachniss, Cyrill},
  title   = {Speeding-Up Robot Exploration by Exploiting Background Information},
  journal = {IEEE Robotics and Automation Letters},
  volume  = {1},
  number  = {2},
  pages   = {716--723},
  year    = {2016},
  doi     = {10.1109/LRA.2016.2520560},
  url     = {https://doi.org/10.1109/LRA.2016.2520560}
}

@inproceedings{strom2015predictive,
  author    = {Perea Str{\"o}m, Daniel and Nenci, Fabrizio and Stachniss, Cyrill},
  title     = {Predictive Exploration Considering Previously Mapped Environments},
  booktitle = {2015 IEEE International Conference on Robotics and Automation ({ICRA})},
  pages     = {2761--2766},
  publisher = {IEEE},
  year      = {2015},
  doi       = {10.1109/ICRA.2015.7139574},
  url       = {https://doi.org/10.1109/ICRA.2015.7139574}
}

@inproceedings{shrestha2019learned,
  author    = {Shrestha, Rakesh and Tian, Fei-Peng and Feng, Wei and Tan, Ping and Vaughan, Richard},
  title     = {Learned Map Prediction for Enhanced Mobile Robot Exploration},
  booktitle = {2019 International Conference on Robotics and Automation ({ICRA})},
  pages     = {1197--1204},
  publisher = {IEEE},
  year      = {2019},
  doi       = {10.1109/ICRA.2019.8793769},
  url       = {https://doi.org/10.1109/ICRA.2019.8793769}
}

@inproceedings{tao2023seer,
  author    = {Tao, Yuezhan and Wu, Yuwei and Li, Beiming and Cladera, Fernando and Zhou, Alex and Thakur, Dinesh and Kumar, Vijay},
  title     = {{SEER}: Safe Efficient Exploration for Aerial Robots Using Learning to Predict Information Gain},
  booktitle = {2023 IEEE International Conference on Robotics and Automation ({ICRA})},
  pages     = {1235--1241},
  publisher = {IEEE},
  year      = {2023},
  doi       = {10.1109/ICRA48891.2023.10160295},
  url       = {https://doi.org/10.1109/ICRA48891.2023.10160295}
}

@article{luperto2020robot,
  author  = {Luperto, Matteo and Antonazzi, Michele and Amigoni, Francesco and Borghese, Nunzio Alberto},
  title   = {Robot Exploration of Indoor Environments Using Incomplete and Inaccurate Prior Knowledge},
  journal = {Robotics and Autonomous Systems},
  volume  = {133},
  pages   = {103622},
  year    = {2020},
  doi     = {10.1016/j.robot.2020.103622},
  url     = {https://doi.org/10.1016/j.robot.2020.103622}
}

@inproceedings{hsu2019learning,
  title={Learning-Based Frequency Estimation Algorithms.},
  author={Hsu, Chen-Yu and Indyk, Piotr and Katabi, Dina and Vakilian, Ali},
  booktitle={International Conference on Learning Representations},
  year={2019}
}

@inproceedings{Mitzenmacher18,
  author       = {Michael Mitzenmacher},
  title        = {A Model for Learned Bloom Filters and Optimizing by Sandwiching},
  booktitle    = {Advances in Neural Information Processing Systems (NeurIPS)},
  pages        = {462--471},
  year         = {2018}
}

@inproceedings{KraskaBCDP18,
  author       = {Tim Kraska and
                  Alex Beutel and
                  Ed H. Chi and
                  Jeffrey Dean and
                  Neoklis Polyzotis},
  title        = {The Case for Learned Index Structures},
  booktitle    = {International Conference on Management of Data (SIGMOD)},
  pages        = {489--504},
  publisher    = {{ACM}},
  year         = {2018}
}

@inproceedings{depavia2024learning,
  title={Learning-based algorithms for graph searching problems},
  author={DePavia, Adela F and Tani, Erasmo and Vakilian, Ali},
  booktitle={International Conference on Artificial Intelligence and Statistics},
  pages={928--936},
  year={2024},
  organization={PMLR}
}

@inproceedings{cabello2024searching,
  title={Searching in Euclidean spaces with predictions},
  author={Cabello, Sergio and Giannopoulos, Panos},
  booktitle={International Workshop on Approximation and Online Algorithms},
  pages={31--45},
  year={2024},
  organization={Springer}
}

@article{DBLP:journals/corr/abs-2609-26789,
  author       = {Romain Cosson and
                  Laurent Massouli{\'{e}}},
  title        = {Polylogarithmic Collective Tree Exploration},
  journal      = {CoRR},
  volume       = {abs/2609.26789},
  year         = {2026},
  url          = {https://doi.org/10.48550/arXiv.2609.26789},
  doi          = {10.48550/ARXIV.2609.26789},
  eprinttype   = {arXiv},
  eprint       = {2609.26789},
  bibsource    = {dblp computer science bibliography, https://dblp.org}
}

\appendix
\section{Assumptions}
\label{appendix:assumption}

\begin{lemma}[Bounded Contraction and Expansion]
    Suppose we have a collection of predictions $\hat n$ with distortion $\mu$. Then, we can spend $O(D)$ rounds and transform it online into a collection $\hat n'$ with distortion $\mu$ such that the expansion and contraction are at least $1$.
\end{lemma}
\begin{proof}
    Send one robot from the root down an arbitrary root-to-leaf path and have it return, which takes at most $2D$ rounds. Let $z$ be the leaf eventually reached. Because $z$ is a leaf, $n_z = 1$, and its prediction $\hat n_z$ is therefore an observed prediction ratio. Abbreviate $\alpha:=\hat n_z$. Calibrating every prediction by setting $\hat n_u'=\hat n_u/\alpha$ ensures that $\mu_e, \mu_c \ge 1$ since $\hat n_z = n_z$ while retaining $\mu$.
\end{proof}

\begin{lemma}[Integral Predictions]
\label{lem:integral-predictions}
Suppose we have a collection of nonnegative predictions $\hat n$ with additive error $E$ and distortion $\mu$. Then, we can transform it online into a collection $\hat n'$ such that $\hat n_u' \in \mathbb Z_+$ for all $u$, with additive error at most $E + 1$ and multiplicative distortion at most $2\mu$.
\end{lemma}

\begin{proof}
We simply define $\hat n_u' = \lceil \max(1, \hat n_u)\rceil$, i.e., round up and take the maximum with $1$.

For the additive error, if $\hat n_u<1$, replacing $\hat n_u$ by $1$ can only decrease its distance from the integer $n_u\geq1$. Otherwise, rounding up changes the prediction by less than one. Hence, the new maximum additive error is at most $E+1$. 

For the multiplicative error, since $\hat n_u' \ge \hat n_u$, clearly the contraction is at most $\mu_c$. For the expansion, if $\hat n_u < n_u$, then $\lceil \max(1, \hat n_u)\rceil/n_u \le \hat n_u /n_u$; otherwise, $\hat n_u' \ge n_u \ge 1$, and $\lceil \max(1, \hat n_u') \rceil \le \hat n_u + 1 \le 2\hat n_u$, so the expansion increases by at most a factor of $2$. The resulting multiplicative distortion is consequently at most $2\mu_e\mu_c=2\mu$.
\end{proof}

\begin{lemma}[Monotone Predictions]
    Suppose we have a collection of nonnegative predictions $\hat n$ with additive error $E$ and distortion $\mu$. Then, we can transform it online into a collection $\hat n'$ with the same error parameters $\E$ and $\mu$ with the additional property that $\hat n'_u \le \hat n'_{p(u)}$ for $u \ne r$.
\end{lemma}
\begin{proof}
    Set $\hat n_r' \gets \hat n_r$. Whenever a new vertex $v \in \ch(u)$ is revealed, set $\hat n'_v \gets \min(\hat n_{u}', \hat n_v)$. We show that one such change does not increase the error, and the lemma follows from induction. Intuitively, decreasing can only worsen under-estimations. However, if that happens, the underestimation factor is already dominated by the parent.

    For the additive error: we have $\hat n'_v - n_v \le \hat n_v - n_v$ and \[n_v - \hat n'_v \le \max(n_v - \hat n_v, n_v - \hat n'_u) \le \max(n_v - \hat n_v, n_u - \hat n'_u),\]
    so the error is the same as when the prediction of $v$ is $\hat n_v$.

    Similarly, for the multiplicative distortion:
    \begin{align*}
        \frac{n_v}{\hat n'_v} &\le \max \left(\frac{n_v}{\hat n'_v}, \frac{n_v}{\hat n'_u}\right) \le \max \left(\frac{n_v}{\hat n'_v}, \frac{n_u}{\hat n'_u}\right),\\
     \frac{\hat n'_v}{n_v} &\le \frac{\hat n_v}{n_v} ,
    \end{align*}
    and the distortion is the same before and after the change.
\end{proof}

\section{Consistency and Robustness}
\label{appendix:consistency-robustness}
\begin{lemma}[Consistency]
\label{lem:perfect-prediction}
    If the predictions are perfect, i.e., $\hat n_u = n_u$ for all $u$, then there is an algorithm that completes the task in $O(n/k + D)$ rounds.
\end{lemma}
\begin{proof}
    If the tree is known, one way to achieve $O(n/k+D)$ is the following. Consider an Eulerian tour of the tree of length $2n-1$. Partition the Eulerian tour into $k$ chunks of length $2n/k$ each. Since consecutive vertices in the Eulerian tour are adjacent in the tree, sending each robot to explore a chunk and return takes $O(n/k+D)$. 
    
    Note that it suffices to know the size of each subtree to simulate such an algorithm, as a subtree of size $s$ induces an interval of length $2s$ in the Eulerian tour. In particular, since $\widehat n_r=n$ is known at time zero, every robot knows its assigned interval. 

To locate the first occurrence of its segment, suppose that the robot is at a vertex $u$. For every child $w$
of $u$, the corresponding excursion in the Eulerian tour contains exactly
$2n_w$ subsequent vertex occurrences. Since the predictions are exact, upon visiting $u$ the robot knows these lengths and can determine which child excursion contains its desired
occurrence. It then moves to that child and repeats, unless the desired
occurrence is the return occurrence of $u$. Each such step moves one
level downward, so the robot locates its starting occurrence, together
with the corresponding depth-first-search state, in at most $D$
rounds. The robot then follows its assigned segment of the tour. This procedure simulates the aforementioned Eulerian tour algorithm and completes the task in $O(n/k+D)$ rounds.
\end{proof}

We remark that the algorithm above does not directly extend to erroneous
predictions. It uses subtree-size predictions as exact excursion
lengths to locate each robot's starting occurrence. With errors,
these lengths need not be consistent between a vertex and its
children, so a prescribed position may correspond to neither a
child excursion nor a return occurrence. Thus the algorithm as
stated need not be well defined.

A natural repair is to clip an invalid position to an available
occurrence. This can make several robots start at the same
occurrence, causing their segments to overlap and leaving other
parts of the tree unexplored. Alternatively, robots could take
over the work left uncovered by these segments. However, a robot
may then inherit unfinished work from several underestimated
subtrees, so the extra work can accumulate as a sum of prediction
errors rather than remain bounded by the maximum error $E$.

\begin{lemma}[Min-operator and Robustness]\label{lem:min}
  The problem of collective tree exploration admits a min-operator, i.e., for two algorithms $\mc A, \mc B$ with completion times $T_{\mc A}, T_{\mc B}$, there is an algorithm with completion time $O(1)\min(T_{\mc A}, T_{\mc B})$.
\end{lemma}
\begin{proof}
  For $i=0,1,2,\ldots$, the algorithm simulates $\mc A$ for $2^i$ steps. If the tree is not fully explored, it returns to the initial configuration and simulates $\mc B$ for $2^i$ steps. The algorithm repeats this procedure until the tree is fully explored. The total cost is at most a constant factor larger than the smaller of the two running times.
\end{proof}

\end{document}